\documentclass[fleqn,usenatbib,onecolumn]{mnras}
\usepackage[T1]{fontenc}
\usepackage{ae,aecompl}
\usepackage{graphicx}	
\usepackage{amsmath}	
\usepackage{amssymb}	
\usepackage{color}
\usepackage[dvipsnames]{xcolor}

\newcommand{\be}{\begin{equation}}
\newcommand{\ee}{\end{equation}}
\newcommand{\bq}{\begin{eqnarray}}
\newcommand{\eq}{\end{eqnarray}}
\title[Redshift drift cosmography]{Redshift drift cosmography with ELT and SKAO measurements}

\author[B. A. R. Rocha and C. J. A. P. Martins]{
B. A. R. Rocha,$^{1,2}$\thanks{E-mail: up201604851@fc.up.pt (BARR)}
and C. J. A. P. Martins$^{1,3}$\thanks{E-mail: Carlos.Martins@astro.up.pt (CJAPM)}\\
$^{1}$Centro de Astrof\'{\i}sica da Universidade do Porto, Rua das Estrelas, 4150-762 Porto, Portugal\\
$^{2}$Faculdade de Ci\^encias, Universidade do Porto, Rua do Campo Alegre, 4150-007 Porto, Portugal\\
$^{3}$Instituto de Astrof\'{\i}sica e Ci\^encias do Espa\c co, CAUP, Rua das Estrelas, 4150-762 Porto, Portugal\\
}

\date{Accepted XXX. Received YYY; in original form ZZZ}
\pubyear{2022}

\begin{document}
\label{firstpage}
\pagerange{\pageref{firstpage}--\pageref{lastpage}}
\maketitle

\begin{abstract}
Mapping the expansion history of the universe is a compelling task of physical cosmology, especially in the context of the observational evidence for the recent acceleration of the universe, which demonstrates that canonical theories of cosmology and particle physics are incomplete and that there is new physics still to be discovered. Cosmography is a phenomenological approach to cosmology, where (with some caveats) physical quantities are expanded as a Taylor series in the cosmological redshift $z$, or analogous parameters such as the rescaled redshift $y=z/(1+z)$ or the logarithmic redshift $x=\ln{(1+z)}$. Moreover, the redshift drift of objects following cosmological expansion provides a model-independent observable, detectable by facilities currently under construction, \textit{viz.} the Extremely Large Telescope and the Square Kilometre Array Observatory (at least in its full configuration). Here we use simulated redshift drift measurements from the two facilities to carry out an assessment of the cosmological impact and model discriminating power of redshift drift cosmography. We find that the combination of measurements from the two facilities can provide a stringent test of the $\Lambda$CDM paradigm, and that overall the logarithmic based expansions of the spectroscopic velocity drift are the most reliable ones, performing better than analogous expansions in the redshift or the rescaled redshift: the former nominally gives the smaller error bars for the cosmographic coefficients but is vulnerable to biases in the higher order terms (in other words, it is only reliable at low redshifts), while the latter always performs poorly.
\end{abstract}



\section{Introduction}
\label{intro}

Modern cosmology is at an exciting stage. There is a standard model, the so-called $\Lambda CDM$ or concordance model, a remarkably simple yet successful way to describe the Universe. At the same time, there is growing observational evidence that it is merely a good phenomenological approximation to a still unknown but more fundamental model. Indeed, the observational evidence for the acceleration of the universe shows that new physics is out there, waiting to be discovered. Current and forthcoming astrophysical facilities must search for, identify and ultimately characterize this new physics.

One important limitation in this quest so far is that the interpretation of all our existing cosmological observations is model-dependent, in the sense that it requires a number of \textit{a priori} assumptions on an underlying cosmological model. Ideally, one would like to rely on fully model-independent observables, enabling consistency tests of these underling assumptions. In principle there are ways to mitigate this current limitation, and in this work we discuss what is arguably the best such example---which is currently not feasible but is within the reach of the next generation of astrophysical facilities.

Indeed, cosmology contains at least one conceptually simple, though operationally very challenging, model-independent observable: the redshift drift of objects following the cosmological expansion, also known as the Sandage test \citep{Sandage,Mcvittie}. This makes no assumption on geometry, clustering or the behaviour of gravity, and is therefore of crucial importance for fundamental cosmology. The practical challenge stems from the fact that cosmologically relevant timescales are orders of magnitude larger than human timescales, and therefore a measurement of the redshift drift requires exquisite experimental sensitivity if it is to be done in manageable amounts of both experiment time and observation time. Current facilities only allow sensitivities about three orders of magnitude worse than the signal expected under standard $\Lambda$CDM assumptions \citep{Darling,Cooke,Lu}. Nevertheless, previous feasibility studies suggest that it is possible to measure the redshift drift with forthcoming facilities, including the Extremely Large Telescope (ELT), as detailed in \citet{Liske}, and the Square Kilometre Array Observatory (SKAO), as detailed in \citet{Klockner}, although for the latter the full (Phase 2) configuration will be necessary. More broadly, redshift drift measurements are a key part of what is commonly called real-time cosmology---see \citet{Quercellini} for a review. All of these pertain to the first derivatives of the redshift, but for the full SKAO one may even foresee measuring second derivatives of the redshift \citep{Second}.

One may envisage different ways to use such redshift drift measurements. A first and more direct one is simply to map the drift rate as a function of redshift. This is fully model-independent but already conveys key cosmological information, since a positive drift rate at low redshifts is a fingerprint of cosmic acceleration. A second one is to use the measurements to constrain specific cosmological models. Several early analyses focused on the ELT measurements \citep{Corasaniti,Balbi,Moraes,CMB}. The important overall result of these works is that although ELT redshift drift measurements, on their own, lead to cosmological parameter constraints that are not tighter than those available by more classical probes (such as type Ia supernovae or the cosmic microwave background) they do probe regions of parameter space that are different from (and sometimes actually orthogonal to) those of other probes, enabling the breaking of degeneracies and therefore leading to more stringent combined constraints. Detailed comparative assessments, including various astrophysical facilities, have been recently reported by \citet{Alves} and \citet{Esteves}.

A third possibility, which we explore in this work, relies on cosmography \citep{Visser}. This is a phenomenological and relatively model-independent approach to cosmology, where physical quantities are expanded as a Taylor series, either in the cosmological redshift $z$ or in related variables, around the present time---that is, zero redshift. This removes the need for the \textit{a priori} choice of a specific model other than, say, assuming that it is of the Friedmann-Lema\^{\i}tre-Robertson-Walker class. This does have some shortcomings, the first of which is that a Taylor series in $z$ might diverge for $z>1$, although this can be mitigated by expanding in other variables, such as a rescaled redshift parameter \citep{Cattoen}. Other limitations include the need to choose where the series is truncated, degeneracies between the series coefficients, and a dependence on chosen priors for these coefficients \citep{Dunsby}. Nevertheless, cosmography provides useful constraints on important quantities such as the deceleration and jerk parameters. A cosmographic expansion of the redshift drift has been previously considered on purely abstract (mathematical) grounds \citep{Math1,Math2}. Here we use simulated redshift drift measurements from the ELT and the SKAO to carry out an assessment of the cosmological impact and model discriminating power of redshift drift cosmography.

The outline of the rest of the paper is as follows. We start in Sect. \ref{formalism} by defining the redshift drift itself and introducing its cosmographic series, in terms of various expansion parameters, all of which have been previously considered for the cosmographic analysis of other (more classical) observables. In addition to the redshift $z$, we also consider the rescaled redshift $y=z/(1+z)$ and the logarithmic redshift $x=\ln{(1+z)}$, as well as expansions based on Pad\'e approximants \citep{Press}. Section \ref{fiducial} defines our fiducial cosmological models, and then provides a detailed assessment of the reliability of the various expansions. One important point is that the relative errors due to the series approximation are different for the Hubble parameter and for the velocity drift (the latter being the actual observable in this case). In Sect. \ref{datasets} we briefly describe the assumptions underlying our simulated SKAO and ELT data and our choices of priors for the series coefficients. Our results are presented in Sect. \ref{results}, where we first separately describe the analysis of the SKAO and ELT data, and then describe the combination of the two data sets. Additionally we consider the case of an ideal data set---although gathering such a data set would have a prohibitive cost in terms of telescope time, this serves the purpose of highlighting the potential as well as the limitations of the cosmographic approach. Finally, our conclusions are in Sect. \ref{concl}.

\section{Mathematical formalism}
\label{formalism}

The redshift drift of an astrophysical object following the cosmological expansion, for an observer which tracks it through a time span $\Delta t$, is given by \citep{Sandage,Liske,Second}
\be
\frac{\Delta z}{\Delta t}=H_0 \left[1+z-E(z)\right]\,,
\ee
although the actual astrophysical observable is a spectroscopically measured velocity drift
\be\label{specvel}
\Delta v=\frac{c\Delta z}{1+z}=(cH_0\Delta t)\left[1-\frac{E(z)}{1+z}\right]\,;
\ee
for future convenience we have defined the rescaled Hubble parameter
\be
E(z)=\frac{H(z)}{H_0}\,,
\ee
with $H_0$ denoting the present-day value of the Hubble parameter (in other words, the Hubble constant). The dependence on the Hubble parameter $H(z)$ naturally leads to a redshift dependence of the drift, which furthermore will be model-dependent. Broadly speaking, in a universe that is currently accelerating but was decelerating in the past the drift will be positive at low redshifts and negative for higher redshifts, while in a universe that always decelerates the redshift drift would always be negative. For the canonical cosmological model, the expected signal is of the order of a few $cm/s$ in a decade.

Much in the same way that redshift drift measurements are model-independent, it is possible to characterize the Universe's evolution without postulating \textit{a priori} a specific model through cosmography, also called cosmokinetics \citep{Visser}. This does rely on the symmetries and geometry of a Friedmann-Lema\'{\i}tre-Robertson-Walker spacetime but it does not use directly Einstein's field equations, thus keeping the number of assumptions to a minimum. Under these assumptions it is possible to write the scale factor as a Taylor expansion around the present time, $t_0$. Defining $t_{H}=H_0(t-t_0)$ for simplicity we have, up to sixth order\footnote{The zero index always denotes the present-day value of the corresponding quantity.}
\begin{equation}\label{scale}
\frac{a(t_{H})}{a_0}=1 + t_{H} -\frac{q_0}{2}(t_{H})^2 + \frac{j_0}{3!}(t_{H})^3 + \frac{s_0}{4!}(t_{H})^4 + \frac{c_0}{5!}(t_{H})^5 + \frac{p_0}{6!}(t_{H})^6 + O((t_{H})^7)\,.
\end{equation}
The cosmographic coefficients are the present-day values of parameters that are respectively denoted the Hubble parameter ($H$), the deceleration parameter ($q$), the jerk ($j$), the snap ($s$), the crackle ($c$) and the pop ($p$)\footnote{The crackle and pop parameters can also be called the lerk $(l$) and the max-out ($m$) parameters, respectively.}, generically defined as
\begin{equation}
H \equiv \frac{1}{a}\frac{da}{dt}\,, \quad \quad q \equiv -\frac{1}{aH^2}\frac{d^2a}{dt^2}\,, \quad \quad j \equiv \frac{1}{aH^3}\frac{d^3a}{dt^3}\,, \quad \quad
s \equiv \frac{1}{aH^4}\frac{d^4a}{dt^4}\,, \quad \quad c \equiv \frac{1}{aH^5}\frac{d^5a}{dt^5}\,, \quad \quad p \equiv \frac{1}{aH^6}\frac{d^6a}{dt^6}\,.
\end{equation}
Each of these parameters encodes information about the expansion history of the universe. Specifically, the sign of the deceleration parameter identifies whether Universe's expansion is accelerating or decelerating---and previous studies suggest that the redshift drift is an optimal way to measure it \citep{Neben}. Similarly, the jerk parameter provides a litmus test for flat $\Lambda$CDM \citep{Second}, and the snap is one way of quantifying how dark energy evolves \citep{Dunsby}. The practical relevance of the higher-order parameters is less clear.

To obtain the cosmographic series for the redshift drift, the main stepping stone is the corresponding series for the Hubble parameter. After some algebra one obtains
\begin{equation}
\begin{split}
\label{hubcosm}
E(z) & = 1 + (q_0 + 1)z + \frac{1}{2}\left(j_0 - q_0^2\right) z^2 + \frac{1}{3!}\left(-3j_0 -4j_0q_0 + 3q_0^3 + 3q_0^2 - s_0\right)z^3 \\ 
& + \frac{1}{4!}\left(c_0 - 4j_0^2 + 25j_0q_0^2 + 32j_0q_0 + 12j_0 - 15q_0^4- 24q_0^3 - 12q_0^2 + 7q_0s_0 + 8s_0\right)z^4 \\ & + \frac{1}{5!}\left(-11c_0q_0 - 15c_0 + 70j_0^2q_0 + 60j_0^2 - 210j_0q_0^3 - 375j_0q_0^2 - 240j_0q_0 + 15j_0s_0 - 60j_0  \right. \\ & \left.- p_0 + 105q_0^5 + 225q_0^4 + 180q_0^3 - 60q_0^2s_0 - 105q_0s_0 + 60q_0^2 - 60s_0\right)z^5 + O(z^6)\,.
\end{split}
\end{equation}
Finally, we can obtain the cosmographic expansion for the redshift drift
\begin{equation}
\begin{split}
\label{dzdt}
\frac{1}{H_0}\frac{\Delta z}{\Delta t} & = -q_0z + \frac{1}{2}\left(q_0^2-j_0\right) z^2 + \frac{1}{3!}\left(3j_0 +4j_0q_0 - 3q_0^3 - 3q_0^2 + s_0\right)z^3 \\ 
& + \frac{1}{4!}\left( 4j_0^2-c_0 - 25j_0q_0^2 - 32j_0q_0 - 12j_0 + 15q_0^4+ 24q_0^3 + 12q_0^2 - 7q_0s_0 - 8s_0\right)z^4 \\ & + \frac{1}{5!}\left(11c_0q_0 + 15c_0 - 70j_0^2q_0 - 60j_0^2 + 210j_0q_0^3 + 375j_0q_0^2 + 240j_0q_0 - 15j_0s_0 + 60j_0  \right. \\ & \left.+ p_0 - 105q_0^5 - 225q_0^4 - 180q_0^3 + 60q_0^2s_0 + 105q_0s_0 - 60q_0^2 + 60s_0\right)z^5 + O(z^6)\,,
\end{split}
\end{equation}
in agreement with what has been found in the literature \citep{Math1}. The same can be done for the observationally more relevant spectroscopic velocity drift
\begin{equation}
\label{dv}
\Delta v = \frac{cH_0\Delta t}{1+z} \bigg[\cdots \bigg]\,,
\end{equation}
where the term inside the square brackets coincides with the right-hand side of Eq. (\ref{dzdt}). Note that these two expressions make it obvious that low-redshift decelerating and accelerating universes will have negative and positive redshift drifts, respectively.

As may be expected from such a simple way to describe the Universe, cosmography faces some shortcomings. The most immediate one is that Taylor series might be expected to diverge for redshifts beyond $z=1$ \citep{Cattoen}, so coefficients that can be easily determined for low redshifts, such as the deceleration and jerk parameters, are expected to be more reliable and easier to study than higher-order ones. One way to mitigate this problem is to replace the redshift by other variables in the Taylor series. A new parameterization will not change the underlying physics but can improve the convergence radius. Other issues are the choice of the order at which the series is truncated and the degeneracies between the cosmographic coefficients---with the exception of the Hubble constant, which is effectively a normalization factor \citep{Dunsby}. A related but equally important issue, which stems from the said degeneracies, is that cosmography can be very dependent on the choices of priors for the cosmographic coefficients.

Regarding the convergence issue, various possible alternative variables have been considered. A recent Masters thesis provides a comparative overview \citep{Thesis}. In what follows we will consider some of these possibilities. The most common alternative variable is the rescaled redshift \citep{Cattoen}, defined as
\be\label{defy}
y\equiv\frac{z}{1+z}\,.
\ee
In this case the cosmographic expansion for the Hubble parameter is
\begin{equation}
\begin{split}
E(y) & = 1 + (1+q_0)y + \frac{1}{2}(j_0-q_0^2+2q_0+2)y^2 - \frac{1}{3!}(4j_0q_0-3j_0-3q_0^3+3q_0^2+s_0-6q_0-6)y^3  \\ &  +\frac{1}{4!}(c_0-4j_0^2+25j_0q_0^2-16j_0q_0+12j_0-15q_0^4+12q_0^3-12q_0^2+7q_0s_0 -4s_0+24q_0+24)y^4 \\ & -\frac{1}{5!}(11c_0q_0-5c_0-70j_0^2q_0 +20j_0^2+210j_0q_0^3-125j_0q_0^2 +80j_0q_0-15j_0s_0-60j_0\\ & +p_0-105q_0^5+75q_0^4-60q_0^3+60q_0^2s_0+60q_0^2-35q_0s_0 +20s_0-120q_0-120)y^5 + O(y^6)\,,
\end{split}
\end{equation}
which again is in agreement with the literature \citep{Xia}. From this we can easily obtain the expressions for the redshift drift and the spectroscopic velocity drift expressed in terms of $y$,
\begin{equation}
\label{redy}
\frac{1}{H_0}\frac{\Delta y}{\Delta t} = (1-y) - (1-y)^2E(y)
\end{equation}
\begin{equation}
\label{specy}
\Delta v = \frac{c\Delta y}{1-y} = c H_0 \Delta t \left[1-(1-y)E(y)\right]\,.
\end{equation}

A second possibility consists of using a logarithm-based expansion \citep{Dunsby,Capozziello}, specifically defining
\begin{equation} \label{xinz}
x = \ln{(1+z)}\,.
\end{equation}
In this case the cosmographic expansion for the Hubble parameter is
\begin{equation}
\begin{split}
E(x) & = 1 + (q_0 + 1)x + \frac{1}{2}\left(j_0 - q_0^2 + q_0 + 1\right) x^2 + \frac{1}{3!}\left(-4j_0q_0 + 3q_0^3 + q_0 - s_0 + 1\right)x^3 \\ 
& + \frac{1}{4!}\left(c_0 - 4j_0^2 + 25j_0q_0^2 + 8j_0q_0 + j_0 - 15q_0^4- 6q_0^3 - q_0^2 + 7q_0s_0 + q_0 + 2s_0 + 1\right)x^4 \\ & + \frac{1}{5!}\left(-11c_0q_0 - 5c_0 + 70j_0^2q_0 + 20j_0^2 - 210j_0q_0^3 - 125j_0q_0^2 - 20j_0q_0 + 15j_0s_0 - p_0 \right. \\ & \left. + 105q_0^5 + 75q_0^4 + 15q_0^3 - 60q_0^2s_0 - 35q_0s_0 + q_0 - 5s_0 + 1\right)x^5 + O(x^6)\,,
\end{split}
\end{equation}
and we can analogously obtain the expressions for the redshift drift and the spectroscopic velocity drift expressed in terms of $x$, the first of which looks particularly simple
\begin{equation}
\frac{1}{H_0}\frac{\Delta x}{\Delta t} =1 - e^{-x}E(x)
\end{equation}
\begin{equation}
\Delta v = c\Delta x = c H_0 \Delta t \left[1-e^{-x}E(x)\right]\,.
\end{equation}

Last but not least, it is also possible to use models based on rational approximations, instead of polynomial approximations constructed when using a Taylor series. An interesting example of this is provided by Pad\'e approximants \citep{Capozziello}. A Pad\'e approximant is a rational function that best fits a given power series $A(x)$ \citep{Press}. The numerator $p(x)$ will be a polynomial with degree $m$ and the denominator $q(x)$ will have degree $n$, such that
\begin{equation}
A(x) = \sum{a_ix^i} = \frac{p(x)}{q(x)} = \frac{p_0+p_1x+p_2x^2+...+p_mx^m}{1+q_1x+q_2x^2+...+q_nx^n}
\end{equation}
The values of $m$ and $n$ added together must be equal to the value of the highest order in the power series. In the present work this will always be equal to five, since  we have redshift terms up to fifth order. The Pad\'e approximants are usually denoted $[x,m/n]$, where the first symbol denotes the variable and the latter two are the degrees of the numerator and denominator polynomials. Thus the approximants $P[x,1/4]$ and $P[x,3/2]$ are
\begin{equation}
\label{padeexample}
P[x,1/4] = \frac{p_0+p_1x}{1+q_1x+q_2x^2+q_3x^3+q_4x^4} \quad \quad \quad P[x,3/2] = \frac{p_0+p_1x+p_2x^2+p_3x^3}{1+q_1x+q_2x^2}
\end{equation}

Again, this alternative approach has its own advantages and disadvantages. Since Pad\'e approximants are rational functions, they are typically less oscillatory than polynomial functions, can fit a wider range of curves and have well-behaved asymptotic properties \citep{Press,Guthrie}. As for drawbacks, there is still no clear way to determine what is the best value to choose for the degrees of the numerator and denominator and sometimes vertical nuisance asymptotes can appear in an unpredictable way \citep{Guthrie}. More details on the expansions based on of Pad\'e approximants can be found in Appendix \ref{AppendixA}.

\section{Fiducial models and series reliability}
\label{fiducial}

It is important to test the reliability of the various choices of cosmographic series introduced in the previous section, by comparison to some plausible fiducial models. In what follows we use the Chevallier-Polarski-Linder (CPL) parametrization as our most generic fiducial model \citep{CPL1,CPL2}, and, in agreement with contemporary observations, we will further assume flat universes. In this case we can write the Friedmann equation as
\be\label{cpl}
E^2(z)=\Omega_m(1+z)^3+\Omega_\phi(1+z)^{3(1+w_0+w_a)}\exp{\left[-\frac{3w_az}{1+z}\right]}\,,
\ee
with $\Omega_m+\Omega_\phi=1$. With the particular choice $(w_0=-1,w_a=0)$ this reduces to the canonical $\Lambda$CDM model.

\begin{figure}
\begin{center}
\includegraphics[width=0.49\columnwidth,keepaspectratio]{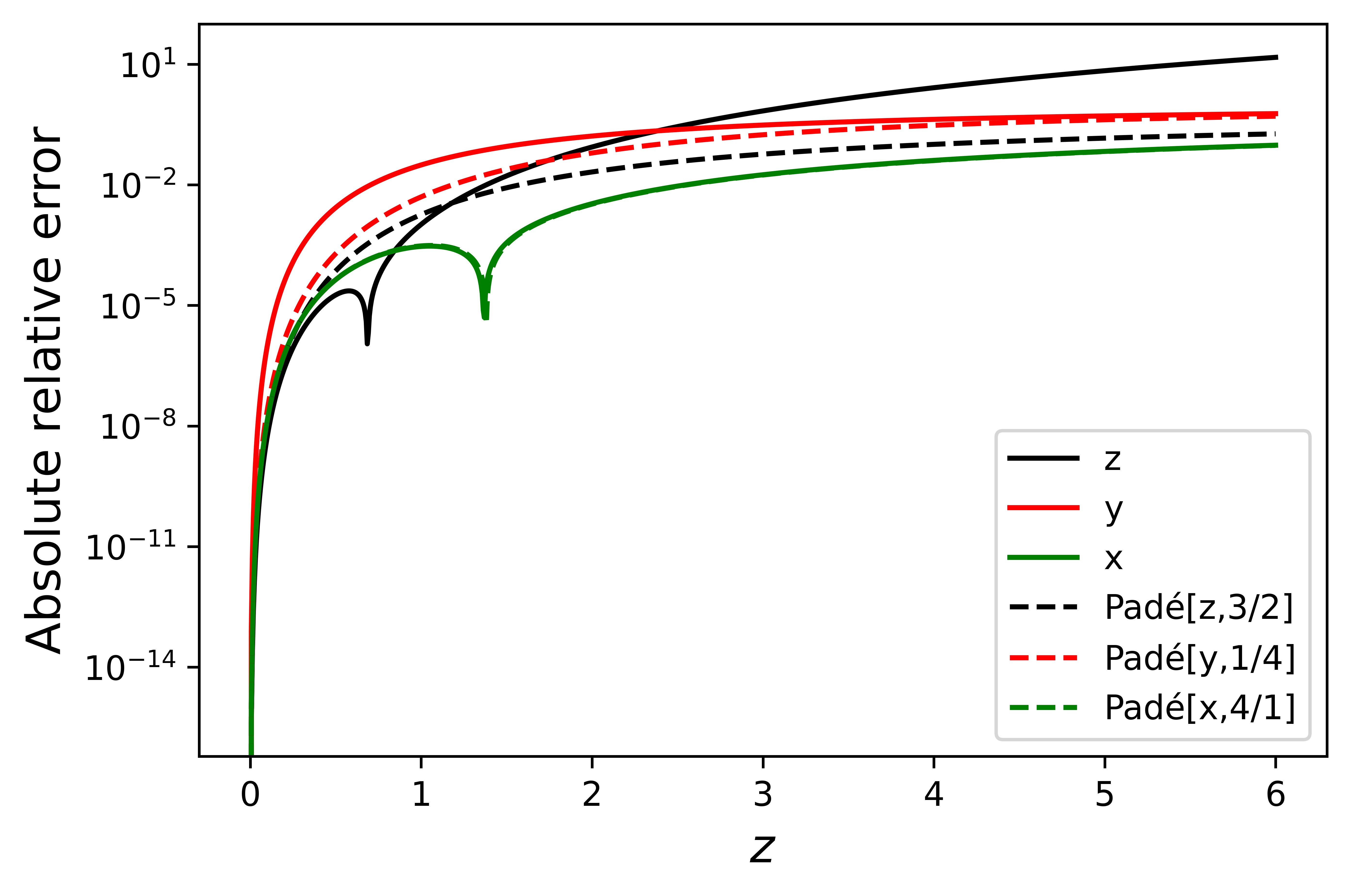}
\includegraphics[width=0.49\columnwidth,keepaspectratio]{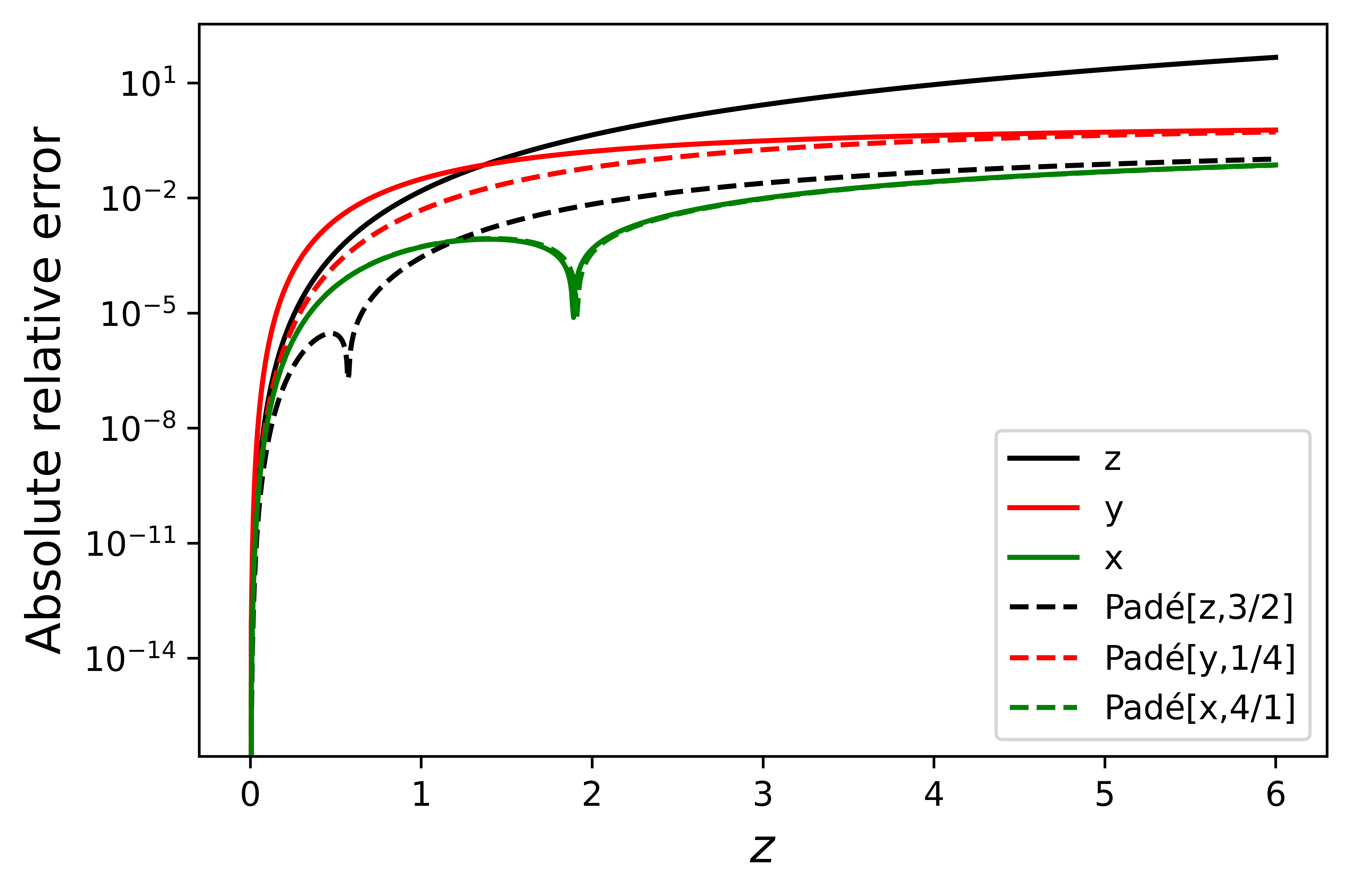}
\end{center}
\caption{Absolute values of the relative error for the cosmographic expansion of the Hubble parameter, $Abs(E_5(z,y,x)-E(z))/E(z)$, where $E_5$ is the relevant cosmographic expansion up to fifth order, for flat fiducial models with $\Omega_{m}=0.3$ and $H_0=70$ km/s/Mpc.The left side panel is for $(w_0=-1,w_a=0)$---\textit{i.e.}, the canonical $\Lambda$CDM model---while the right side panel is for a model with $(w_0=-0.95,w_a=-0.05)$. The solid lines depict the traditional parameterizations and the dashed lines the Pad\'e ones.}
\label{fig01}
\end{figure}

To carry out this comparison, we must know the hyper-parameters of the cosmographic model, \textit{i.e.} the set of values ($H_0$, $q_0$, $j_0$, $s_0$, $c_0$, $p_0$), for each fiducial model, again bearing in mind that $H_0$ only acts as a multiplicative factor in the redshift drift. We start with the simplest case of $\Lambda$CDM, for which the cosmographic coefficients are
\begin{equation}
\label{q0lcdm}
q_0 = -1 + \frac{3}{2}\Omega_{m}\,, \quad j_0 = 1\,, \quad s_0 = 1 - \frac{9}{2}\Omega_{m}\,, \quad c_0 = 1 + 3\Omega_{m} + \frac{27}{2}\Omega_{m}^2\,, \quad p_0 = 1 - \frac{27}{2}\Omega_{m} - 81\Omega_{m}^2 - \frac{81}{4}\Omega_{m}^3
\end{equation}
Again, we note that an observational measurement finding $j_0\neq1$ would immediately disprove the flat $\Lambda$CDM model. Conversely, assuming that this model is correct one can determine the present day matter density parameter from a sufficiently good determination of $q_0$.

\begin{table}
\centering
\caption{Absolute values of the relative error for the cosmographic expansion of the Hubble parameter, $Abs(E_5(z,y,x)-E(z))/E(z)$, where $E_5$ is the relevant cosmographic expansion up to fifth order (left part of table) and the corresponding spectroscopic velocity drift (right part of the table) for the various cosmographic expansions of a fiducial $\Lambda$CDM model with $\Omega_{m0}=0.3$ and $H_0=70$ km/s/Mpc. The errors are listed for some selected values of the redshift. The values for the Pad\'e[x,4/1] are not explicitly listed because they coincide, up to the listed number of significant digits, with those from the regular $x$ parameterization.} \label{tab1}
\begin{tabular}{ |c|c|c|c|c|c|c|c|c|c|c|c|c|  }
    \hline
    \multicolumn{6}{|c|}{Relative error in Hubble parameter} & \multicolumn{1}{|c|}{Cosmographic series} & \multicolumn{6}{|c|}{Relative error in velocity drift} \\
    $z=1$ & $z=2$ & $z=3$ & $z=4$ & $z=5$ & $z=6$ & {} & $z=1$ & $z=2$ & $z=3$ & $z=4$ & $z=5$ & $z=6$ \\
    \hline
    $0.1\%$ & $9\%$ & $69\%$ & $260\%$ & $687\%$ & $1479\%$ & z & $0.8\%$ & $770\%$ & $665\%$ & $1359\%$ & $2658\%$ & $4735\%$ \\
    $0.2\%$ & $2\%$ & $6\%$ & $10\%$ & $14\%$ & $18\%$ & Pad\'e[z,3/2] & $1\%$ & $184\%$ & $56\%$ & $53\%$ & $56\%$ & $59\%$ \\
    \hline
    $3\%$ & $16\%$ & $31\%$ & $42\%$ & $52\%$ & $59\%$ & y & $79\%$ & $4139\%$ & $870\%$ & $694\%$ & $668\%$ & $674\%$ \\
    $0.5\%$ & $6\%$ & $18\%$ & $30\%$ & $41\%$ & $51\%$ & Pad\'e[y,1/4] & $19\%$ & $1296\%$ & $251\%$ & $183\%$ & $162\%$ & $153\%$ \\
    \hline
    $0.03\%$ & $0.3\%$ & $2\%$ & $4\%$ & $7\%$ & $10\%$ & x, Pad\'e[x,4/1] & $0.2\%$ & $30\%$ & $17\%$ & $21\%$ & $26\%$ & $31\%$ \\
    \hline
\end{tabular}
\end{table}

We tested the fitness of the cosmographic expansions for the expansions in $z$, $y$, $x$ and the best performing Pad\'e approximants for each of them, respectively $P[z,3/2]$, $P[y,1/4]$ and $P[x,4/1]$. We assumed a matter density $\Omega_m=0.3$ and a Hubble constant $H_0=70$ km/s/Mpc, which correspond to the following parameters; (70,-0.55,1,-0.35,3.115,-10.88675).

Since the redshift drift is inherently dependent on the Hubble parameter, the relative error obtained from cosmographic approximations for that parameter is the starting point to understand the possible bias present in those expansions. The left-hand side of Figure \ref{fig01} depicts the absolute values of the relative errors for the Hubble parameter in the range $0<z<6$, while the left-hand side of Table \ref{tab1} lists the absolute values of the relative errors for selected redshifts. As expected, all the studied functions are well suited to describe the Hubble parameter for redshifts below $z=1$. The $z$ parameterization has the largest relative error for high redshifts, which is also fully expected. The other two standard parameterizations perform better for higher redshifts, with the logarithmic one being an all around better approximation. The use of Pad\'e approximants dramatically reduces the relative errors at high redshift for the $z$ series, while for the $y$ series this reduction is rather modest, and for the $x$ series the standard Pad\'e series have almost exactly the same relative errors.

On the other hand, the right-hand side of Table \ref{tab1} lists the absolute values of the relative errors, at the same redshifts, for the spectroscopic velocity drift $\Delta v$. This quantity, which is the directly measured one, has a cosmographic series that is related to but nevertheless different from that of the Hubble parameter, and indeed one sees that it is very sensitive to error propagation, with very large relative errors beyond redshift $z=1$. With that caveat, the overall behaviour is similar to that of the previous paragraph: the Pad\'e approximants do better than the traditional expansions for the $z$ and $y$ series, but have a negligible impact in the logarithmic ($x$) expansion---which does significantly better than the others. Clearly one must build upon an extremely good approximation for the Hubble parameter in order to have a reliable redshift drift cosmography at high redshifts.

By way of comparison, we can also obtain the cosmographic coefficients for the full CPL model,
\begin{equation}
q_0 = \frac{1}{2}(w_0(3 - 3\Omega_{m}) + 1)
\end{equation}
\begin{equation}
\label{j0}
j_0 = \frac{1}{2}(w_a(3 - 3\Omega_{m}) + w_0^2(9 - 9\Omega_{m}) + w_0(9 - 9\Omega_{m}) + 2)
\end{equation}
\begin{equation}
\begin{split}
s_0  & = \frac{1}{4}(w_a(33\Omega_{m} - 33) + w_0^3(-27\Omega_{m}^2 + 108\Omega_{m} - 81) + w_0^2(-27\Omega_{m}^2 + 171\Omega_{m} - 144) \\ & + w_0(81\Omega_{m} + w_a(-9\Omega_{m}^2 + 72\Omega_{m} - 63) - 81) - 14)
\end{split}
\end{equation}
\begin{equation}
\begin{split}
c_0  & = \frac{1}{4}(w_a^2(9\Omega_{m}^2 - 72\Omega_{m} + 63) + w_a(213 - 213\Omega_{m}) + w_0^4(324\Omega_{m}^2   - 810\Omega_{m} + 486)+ \\ & w_0^3(648\Omega_{m}^2 - 1917\Omega_{m} + 1269)  + w_0^2(378\Omega_{m}^2 - 1584\Omega_{m} + w_a(297\Omega_{m}^2 - 918\Omega_{m} + 621) \\ & + 1206)  + w_0(-489\Omega_{m} + w_a(189\Omega_{m}^2 - 927\Omega_{m} + 738) + 489) + 70)
\end{split}
\end{equation}
\begin{equation}
\begin{split}
p_0  & = \frac{1}{8}(w_a^2(-459\Omega_{m}^2 + 2502\Omega_{m} - 2043) + w_a(3321\Omega_{m} - 3321) + w_0^5(972\Omega_{m}^3 - 8262\Omega_{m}^2 + 14580\Omega_{m} - 7290) \\ & + w_0^4(1944\Omega_{m}^3 - 23409\Omega_{m}^2 + 46818\Omega_{m} - 25353) + w_0^3(1134\Omega_{m}^3 - 23814\Omega_{m}^2 + 57267\Omega_{m} + w_a(891\Omega_{m}^3 - 11745\Omega_{m}^2 \\ & + 24057\Omega_{m} - 13203) - 34587) + w_0^2(-9315\Omega_{m}^2 + 32328\Omega_{m} + w_a(567\Omega_{m}^3 - 16065\Omega_{m}^2 + 41013\Omega_{m} - 25515) - 23013) \\ &  + w_0(7407\Omega_{m} + w_a^2(27\Omega_{m}^3 - 1863\Omega_{m}^2 + 5265\Omega_{m} - 3429) + w_a(-5508\Omega_{m}^2 + 21645\Omega_{m} - 16137) - 7407) - 910)\,.
\end{split}
\end{equation}
It is straightforward to check that these reduce to the previously introduced expressions for $\Lambda$CDM if one chooses $w_a=0$ and $w_0=-1$. They also agree with the lower order expansions found in the literature \citep{Xia}. 

With a little algebra one can also invert the above expressions and obtain the CPL parameters as a function of the cosmographic ones. One finds
\begin{equation}
\Omega_{m} = \frac{8j_0q_0 + 7j_0 - 16q_0^2 + 2q_0 + 2s_0 - 1 + (1-2q_0)\sqrt{9j_0^2 + 8j_0q_0^2 - 12j_0q_0 - 14j_0 + 32q_0^2 + 8q_0s_0 - 4s_0 + 1}}{2j_0q_0 + 10j_0 - 8q_0 + 2s_0}
\end{equation}
\begin{equation}
w_0 = \frac{1-2q_0}{3(\Omega_{m}-1)}
\end{equation}
\begin{equation}
w_a = \frac{(-2j_0+6q_0-1)\Omega_{m} + 2j_0-4q_0^2-2q_0}{3(\Omega_{m}-1)^2}\,.
\end{equation}
Note that for the flat CPL model to be fully specified one only needs to know the cosmographic parameters $q_0$, $j_0$ and $s_0$. 

A similar analysis can now be done for other fiducial model choices. We keep the same fiducial values for the matter density and the Hubble constant, but now take a dynamical dark energy model, with $(w_0=-0.95,w_a=-0.05)$. The results, which can be visualised in the right-hand side panel of Figure \ref{fig01}, are qualitatively the same, although there is some dependence of the chosen fiducial model parameters. Again the logarithmic ($x$) parametrization is the best behaved, the standard $z$ series gives the poorest results at high redshift, and Pad\'e approximants generally do better than canonical expansions---the difference in relative errors can be up to two orders of magnitude for specific choices of fiducial parameter values, although it is usually smaller than this. We also find that the logarithmic-based ones, \textit{i.e.} $x$ and Pad\'e$[x,4/1]$, only have very similar behaviours for CPL values close to the $\Lambda$CDM; as we move away from this point in parameter space, the Pad\'e$[x,4/1]$ does better than the plain $x$, as is the case for the other variables. A further curiosity is that the mean relative errors of the $y$ parameterization are the least sensitive to the fiducial values of the CPL parameters chosen; indeed in the neighbourhood of the $\Lambda$CDM values they are independent of these parameters.

Given these results, in the rest of the paper we will not further consider the expansions based on the rescaled redshift $y$, but will continue with the three others: the standard and Pad\'e[z,3/2] redshift series, and the logarithmic redshift series---the results for Pad\'e[x,4/1] being similar to the latter, since observations show that any viable cosmological model must be close to $\Lambda$CDM.

\section{Mock data sets}
\label{datasets}

We are now ready to generate our simulated data sets, representative of the redshift drift measurements expected from astrophysical facilities becoming operational in the near future. In most of what follows (and unless otherwise is stated) our fiducial model is the flat $\Lambda$CDM model, with $\Omega_m=0.3$, $H_0=70$ km/s/Mpc, but we will also consider a alternative CPL models.

We emphasize that in generating all these data sets, the simulated measurements were not perturbed by the addition of noise. This is of course an ideal situation, but in the context of our analysis it has the advantage of enabling the identification of possible biases on the cosmographic approach---which, as we shall see, do occur in a few cases. The case with noise has been briefly discussed elsewhere \citep{Thesis}, with the result being the expected one: the errors on the parameters are somewhat increases throughout, without impacting the relative merits of the various cosmographic expansions.

For the ELT, a detailed study has been done by \citet{Liske}, who found that the spectroscopic velocity uncertainty is well approximated by
\be
\sigma_v=1.35\left(\frac{S/N}{2370}\right)^{-1}\left(\frac{N_{QSO}}{30}\right)^{-1/2}\left(\frac{1+z_{QSO}}{5}\right)^{-\lambda} cm/s\,,
\ee
where $S/N$ is the spectral signal to noise ratio per 0.0125 \r{A} pixel, $N_{QSO}$ is the number of objects observed at redshift $z_{QSO}$, and where the last exponent is $\lambda=1.7$ up to $z=4$ and $\lambda=0.9$ for $z>4$. Consistently with this work and other recent detailed studies \citep{Alves,Esteves}, we make the assumption of $N_{QSO}=6$ quasars measured at each of five redshift bins $z=2.0, 2.5, 3.0, 3.5, 4.5$ (i.e., a total sample of 30 quasars) with a time span of $\Delta t=20$ years, and each of the measurement spectra having $S/N=3000$. Although costly in terms of telescope time, these choices, including that of the time span, are consistent with the top-level requirements for the ELT-HIRES\footnote{This is its previous placeholder name, which we use for ease of comparison with the previous literature. The instrument, which is starting its Phase B of construction, is now called ANDES.} spectrograph \citep{HIRES}. The main bottleneck to these measurements (apart from the stability of the spectrograph, which is presently understood not to be a limiting factor) is the availability of sufficiently bright high-redshift quasars, able to deliver the required signal to noise in reasonable amounts of telescope time, but significant progress is being made in this direction. For example, \citet{Boutsia} have recently proposed a possible 'Golden Sample' of 30 such quasars, although the redshift distribution of this sample does not coincide with ours. Meanwhile, other studies suggest using a much smaller number of quasars \citep{Dong}.

For the SKAO, existing studies of the redshift drift science case are much less detailed. Nevertheless we will take these at face value, and specifically we follow \citet{Klockner} and \citet{Second} in assuming ten measurements, equally spaced in redshift between $z=0.1$ and $z=1.0$ with spectroscopic velocity uncertainties correspondingly ranging from $1\%$ at the lowest redshift to $10\%$ at the highest redshift. Defining the data set in terms of relative uncertainties has the advantage of making its implications independent of the experiment's time span and the required telescope time. In other words, optimizing these two specific times is an issue that can be deferred to when the SKAO hardware configuration and other technical specifications are known, and will not directly affect our results. Nevertheless, we note that the work of \citet{Klockner} clearly shows that SKAO redshift drift measurements are only within the reach of its full configuration (formerly known as SKA Phase 2): a redshift drift measurement is unfeasible with the earlier (formerly called SKA Phase 1) configuration, as it would require a time span of about 40 years of data collection. 

Our analysis used a Markov Chain Monte Carlo (MCMC) code, \textit{emcee} for \textit{Python} \citep{Foreman_Mackey_2013}. For the visualization of the MCMC results we used the \textit{Python}'s module \textit{corner} \citep{corner}.

\begin{table}
\centering
\caption{Cosmological priors for the cosmographic coefficients used in our MCMC analysis.} \label{tab2}
\begin{tabular}{ |c|c||c|c| }
    \hline
Parameter name & Symbol & Uniform $H_0$ prior & Normal $H_0$ prior \\
    \hline
Hubble constant &  $H_0$ & $[60,80]$ & $\mathcal{N}(70,3^{2})$ \\
deceleration parameter & $q_0$ & $[-1,1]$ & $[-1,1]$ \\
Jerk & $j_0$ & $[-10,10]$ & $[-10,10]$ \\
Snap & $s_0$ & $[-30,30]$ & $[-30,30]$ \\
Crackle & $c_0$ & $[-50,50]$ & $[-50,50]$ \\
Pop & $p_0$ & $[-100,100]$ & $[-100,100]$ \\
    \hline
\end{tabular}
\end{table}

Last but not least, one needs to define cosmological priors for the cosmographic parameters. We generically use ample uniform (uninformative) priors, shown in Table \ref{tab2}. We note that these are wider than those of most previous cosmographic studies in the literature. For the Hubble constant we consider two scenarios, the first one being a uniform prior with values in the range $[60,80]$ km/s/Mpc and the second one a normal distribution with an average and standard deviation equal to 70 and 3 km/s/Mpc respectively. This choice is meant to span the range of the so-called Hubble tension \citep{Tension}. For the higher-order parameters the priors are agnostically centered at zero (and not at the fiducial value, which will differ in the various cases to be considered), but we have checked that they are always wide enough to have no impact on the results.

\section{Results}
\label{results}

We now discuss the results of our analysis for the previously introduced fiducial models and data sets. We recall that given our findings in Sect. \ref{fiducial} we will only report results for the $z$, Pad\'e$[z,3/2]$ and $x$ series. We first consider the SKAO and ELT data separately, and then discuss the combination of the two data sets. Additional results for other assumptions can be found in \citet{Thesis}.

\subsection{Low-redshift Cosmography: The SKAO}

\begin{figure}
\begin{center}
\includegraphics[width=0.45\columnwidth,keepaspectratio]{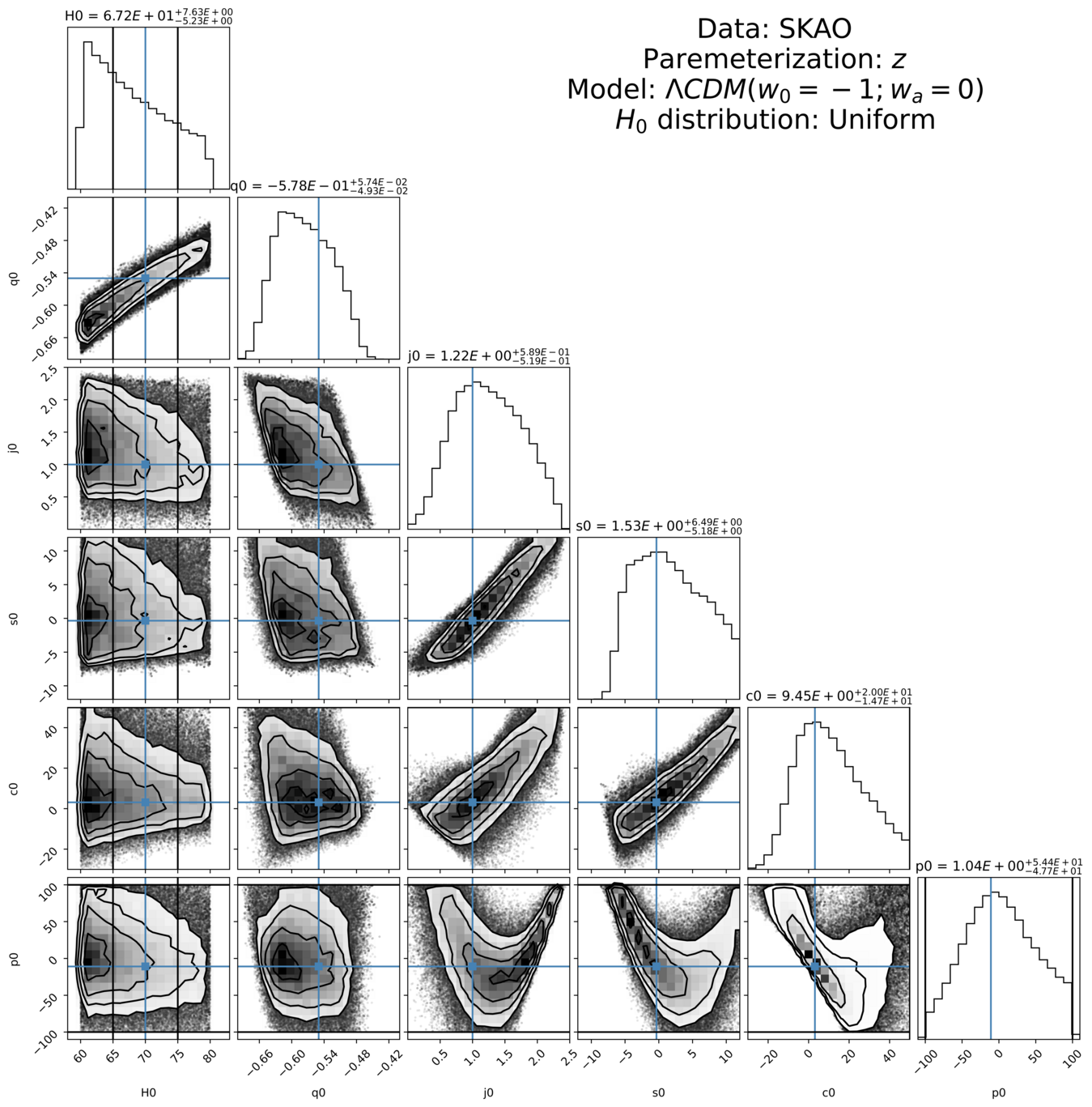}
\includegraphics[width=0.45\columnwidth,keepaspectratio]{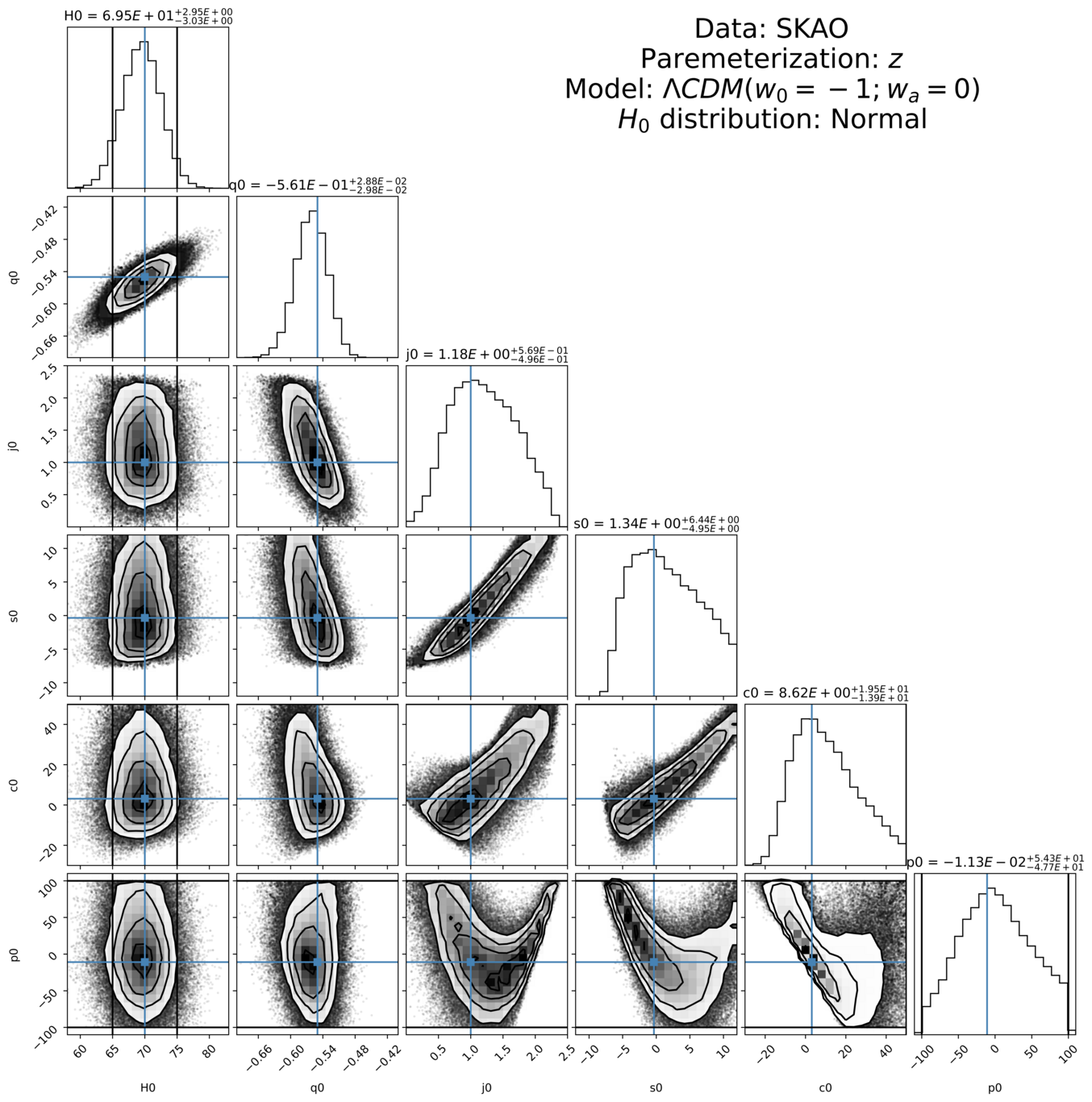}
\includegraphics[width=0.45\columnwidth,keepaspectratio]{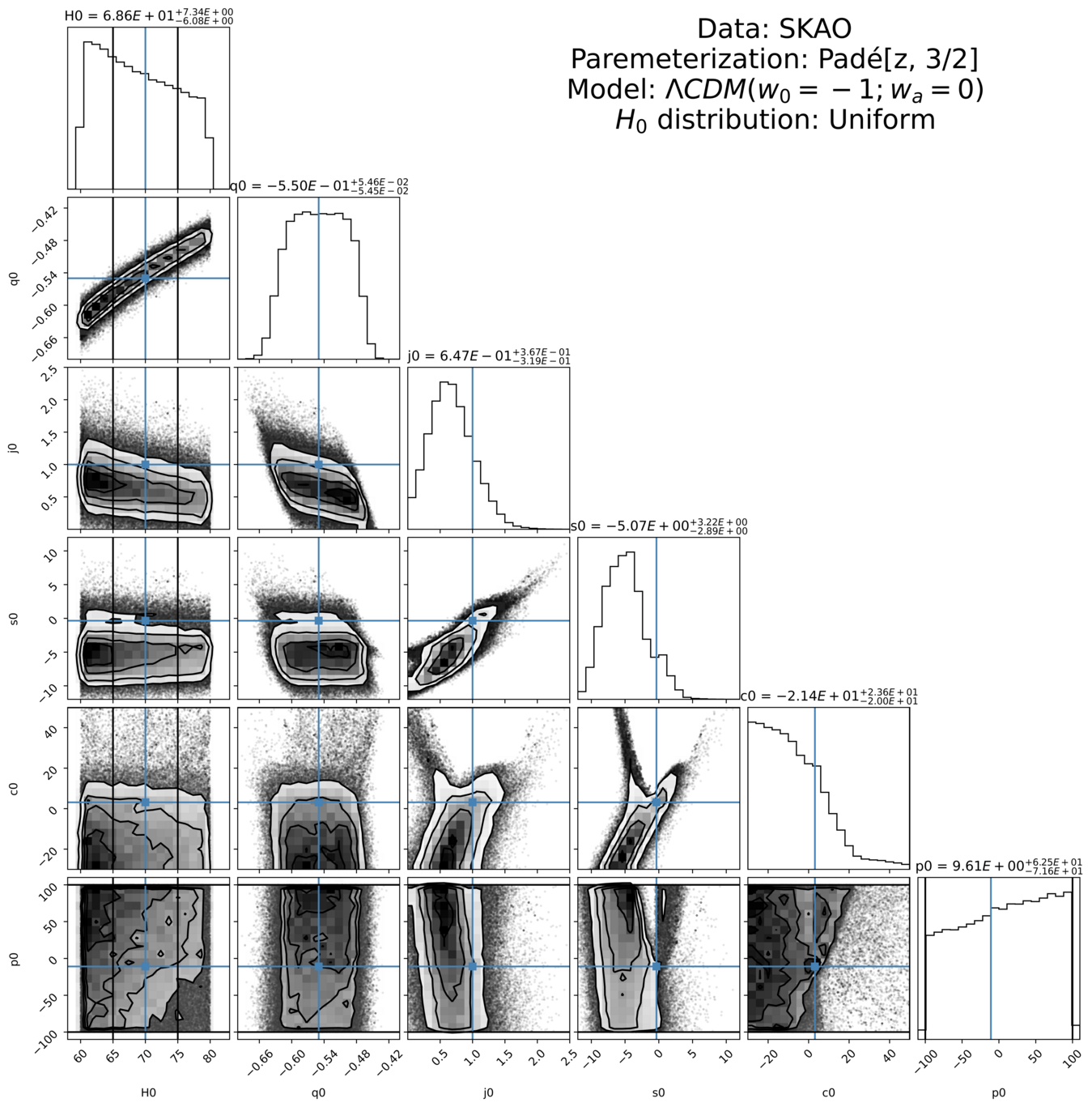}
\includegraphics[width=0.45\columnwidth,keepaspectratio]{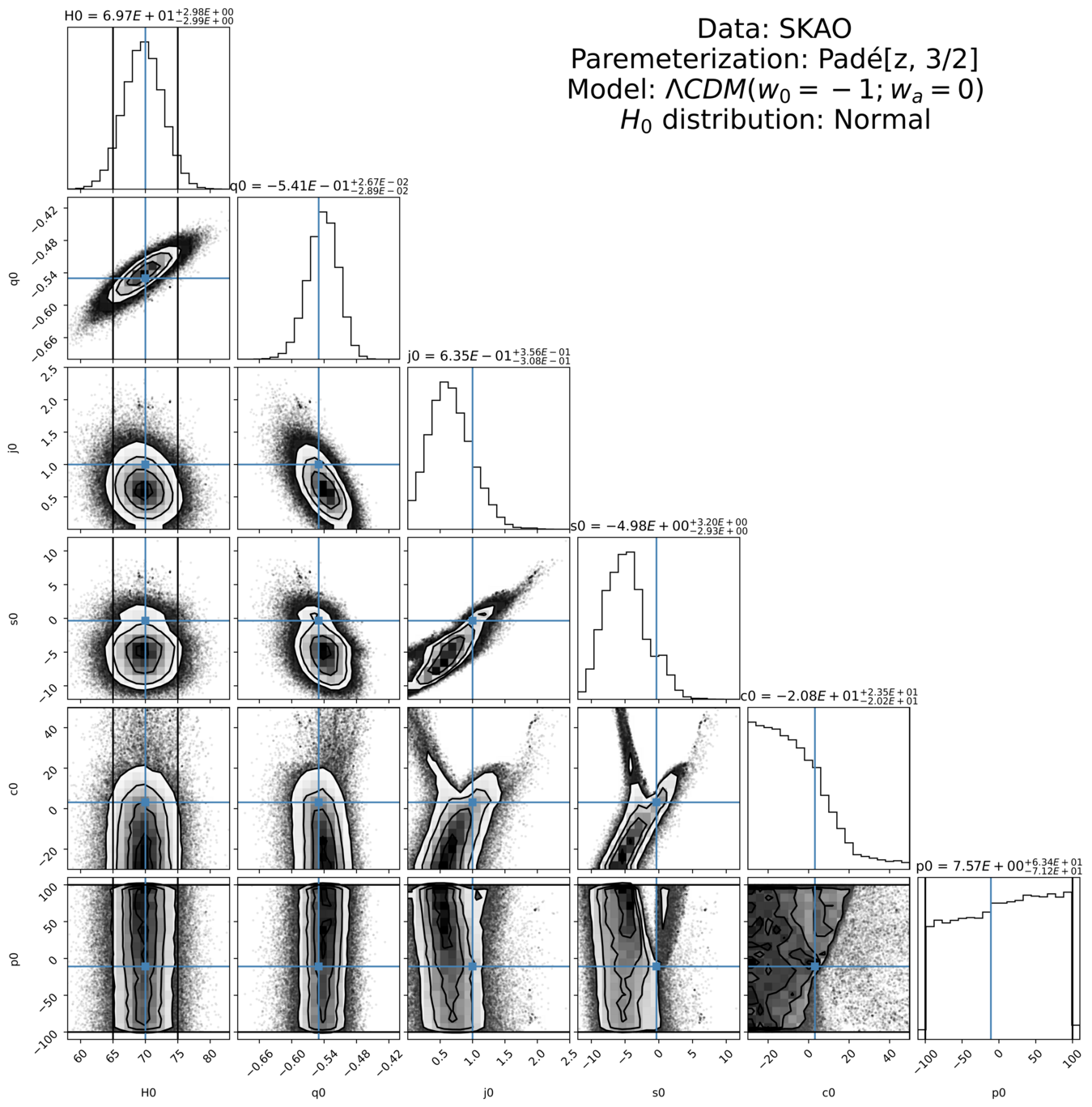}
\includegraphics[width=0.45\columnwidth,keepaspectratio]{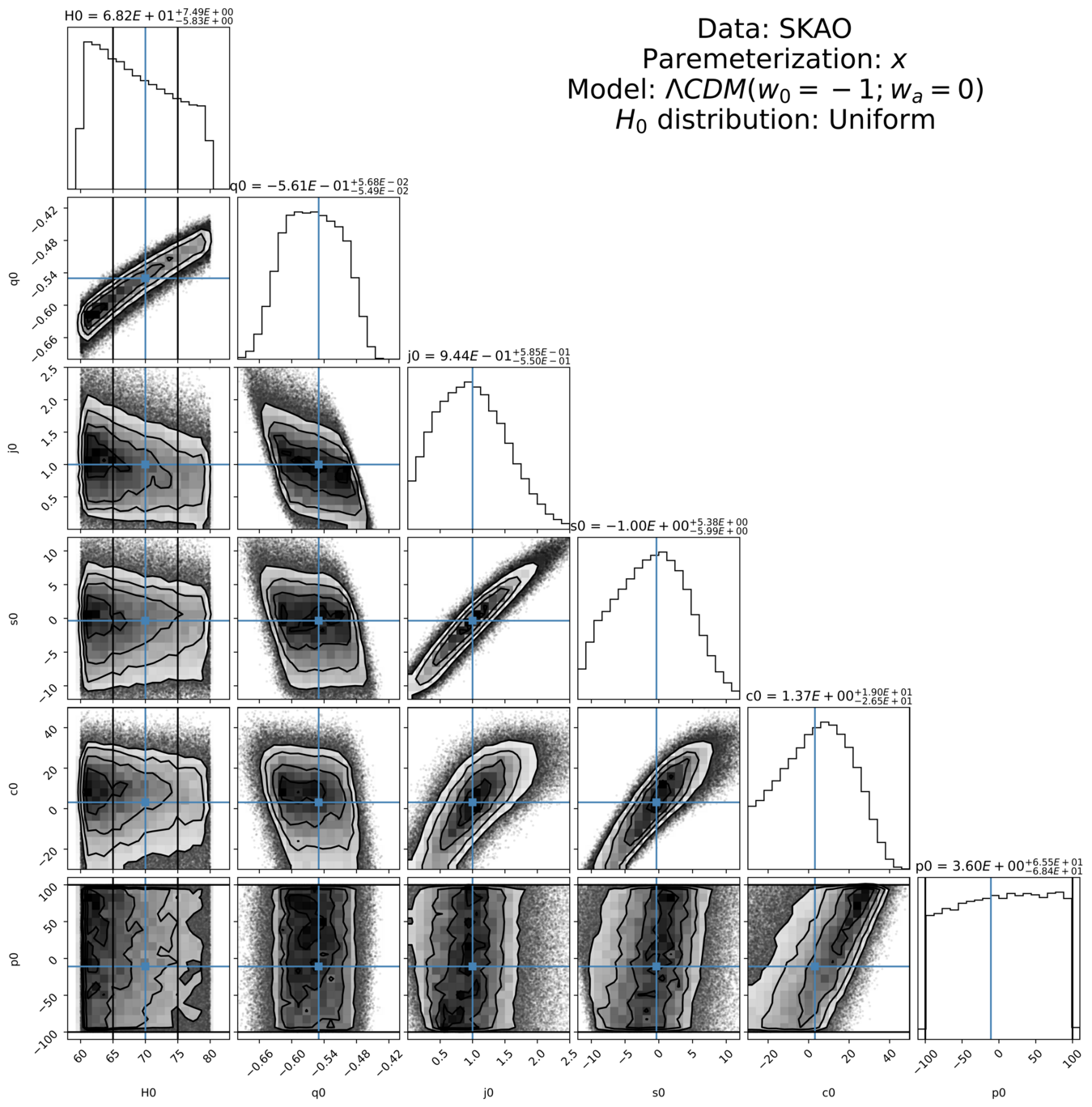}
\includegraphics[width=0.45\columnwidth,keepaspectratio]{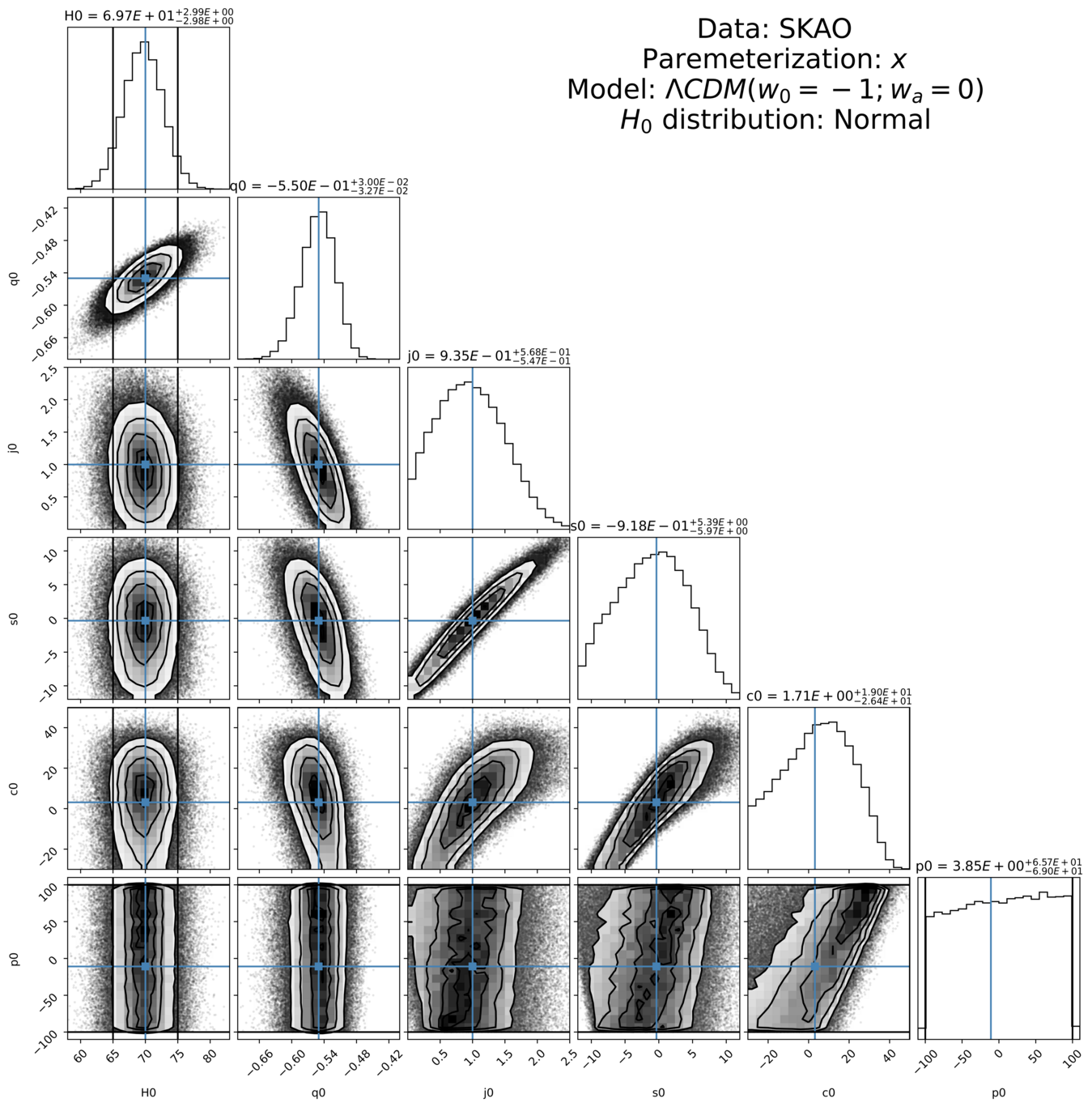}
\end{center}
\caption{Cosmographic parameters from SKAO data with the $z$, Pad\'e$[z,3/2]$ and $x$ series (top, middle and bottom panels). Left and right side panels correspond to uniform and normal priors on $H_0$. The blue lines identify the true (fiducial model) values for each parameter. For $H_0$, the black lines identify the  range of values between 65 and 75 km/s/Mpc; for the other parameters, the black lines (when visible) identify the limits of our priors.}
\label{fig02}
\end{figure}

We start with a comparison of the three cosmographic series for the SKAO data. Figure \ref{fig02} shows the results for the cases of uniform and normal priors on the Hubble constant, and Table \ref{tab3} summarizes the derived parameter constraints for both analyses.

\begin{table}
\centering
\caption{Constraints on the cosmographic coefficients of the three series expansions, obtained from the MCMC analysis of the SKAO data set. The left and right parts of the table correspond to the choices of uniform or normal priors for the Hubble constant. The middle column shows the correct values (approximated to two decimal places) for the assumed fiducial model.}\label{tab3}
\begin{tabular}{ |c|c|c|c|c|c|c|  }
    \hline
     \multicolumn{3}{|c|}{Results for Uniform Prior} & \multicolumn{1}{|c|}{Parameter \& Fiducial value} & \multicolumn{3}{|c|}{Results for Normal Prior} \\
    $z$ & $P[z,3/2]$ & $x$ & { } & $z$ & $P[z,3/2]$ & $x$ \\
    \hline
    $67.21^{+7.63}_{-5.23}$ & $68.58^{+7.34}_{-6.08}$ & $68.24^{+7.49}_{-5.83}$ & $H_0=70$ & $69.51^{+2.95}_{-3.03}$ & $69.70^{+2.98}_{-2.99}$ & $69.69^{+2.99}_{-2.98}$ \\
    $-0.58^{+0.06}_{-0.05}$ & $-0.55^{+0.05}_{-0.05}$ & $-0.56^{+0.06}_{-0.05}$ & $q_0=-0.55$ & $-0.56^{+0.03}_{-0.03}$ & $-0.54^{+0.03}_{-0.03}$ & $-0.55^{+0.03}_{-0.03}$ \\
    $1.22^{+0.59}_{-0.52}$ & $0.65^{+0.37}_{-0.32}$ & $0.94^{+0.58}_{-0.55}$ & $j_0=1$ & $1.18^{+0.57}_{-0.50}$ & $0.63^{+0.36}_{-0.31}$ & $0.93^{+0.57}_{-0.55}$ \\
    $1.53^{+6.49}_{-5.18}$ & $-5.07^{+3.22}_{-2.89}$ & $-1.00^{+5.38}_{-5.99}$ & $s_0=-0.35$ & $1.34^{+6.44}_{-4.95}$ & $-4.98^{+3.20}_{-2.93}$ & $-0.92^{+5.39}_{-5.97}$ \\
    $9.45^{+19.96}_{-14.68}$ & $-21.36^{+23.63}_{-19.98}$ & $1.37^{+19.03}_{-26.51}$ & $c_0=3.12$ & $8.62^{+19.46}_{-13.90}$ & $-20.78^{+23.49}_{-20.23}$ & $1.71^{+18.98}_{-26.35}$ \\
    $1.04^{+54.41}_{-47.70}$ & $9.61^{+62.50}_{-71.63}$ & $3.60^{+65.50}_{-68.42}$ & $p_0=-10.89$ & $-0.01^{+54.30}_{-47.68}$ & $7.57^{+63.38}_{-71.22}$ & $3.85^{+65.70}_{-68.95}$ \\
    \hline
\end{tabular}
\end{table}

We find that all the tested expansions present comparable uncertainties in the constraints for the lower-order cosmographic parameters beyond $H_0$, although it is worthy of note that in the $P[z,3/2]$ expansion the inferred $s_0$ and $c_0$ parameters are biased (by which we mean that the one-sigma posterior constraint does not recover the parameter's fiducial value). With a normal prior on $H_0$ this is also the case for $j_0$. In all cases the deceleration and jerk parameters are accurately determined, with very good precision on the former, and still reasonably precision on the latter (though with the previously mentioned caveat for the $P[z,3/2]$ case). 

On the other hand, for the higher-order parameters the uncertainties are considerably larger, and $p_0$ is almost unconstrained (the magnitude of the one-sigma uncertainty is a large fraction of the prior range). This is of course to be expected, considering that the data stops at redshift $z=1$. As for the impact of the choice of priors on the Hubble constant, while there is no significant reduction in the uncertainties obtained for the cosmographic parameters, the normal prior captures the true values somewhat more accurately than the uniform prior, which is again expected (more on this in the following paragraph). For this reason, in what follows we will simplify our analysis by restricting it to the normal prior case.

Looking at the various panels, one notices several interesting and non-trivial degeneracies. A somewhat surprising one is that between $H_0$ and $q_0$, which is related to the fact that the linear term in the cosmographic series has the form $-H_0q_0z$ (or analogous forms for the other series). As we will see throughout this section, depending on the available data and other choices (including that of the priors) this degeneracy can be reduced, but is never completely eliminated. In this regard, a first important result is that a uniform prior for $H_0$ is much worse than a normal one. This is not at all surprising, since the Hubble constant is a multiplicative factor in the mathematical expression for the redshift drift. Figure \ref{fig02} also shows degeneracies of $s_0$ and $c_0$ with each other and also with $j_0$, but comparison with the results of the following subsections show that these are not generic, in the sense that they can disappear when higher-redshift data is added.

Overall, it is also interesting to note that the simpler $z$ expansion yields lower uncertainties for the higher-order $c_0$ and $p_0$ values, although the difference is perhaps not so relevant since all these uncertainties are quite large. Conversely, the Pad\'e$[z,3/2]$ expansion appears to be worse than the others at recovering true values in the low redshift regime. While we have not explored in detail the reasons for this, the most likely explanation is again the possibility of nuisance asymptotes \citep{Guthrie}, which are commensurate with the peculiarly shaped distributions in some of the two-dimensional panels in Fig. \ref{fig02}. Therefore, this suggests that as long as one only has redshift drift measurements at low redshift, the simplest expansion in redshift works reasonably well, and there is no reason to further complicate it.

\subsection{High-redshift Cosmography: The ELT}

\begin{figure}
\begin{center}
\includegraphics[width=0.45\columnwidth,keepaspectratio]{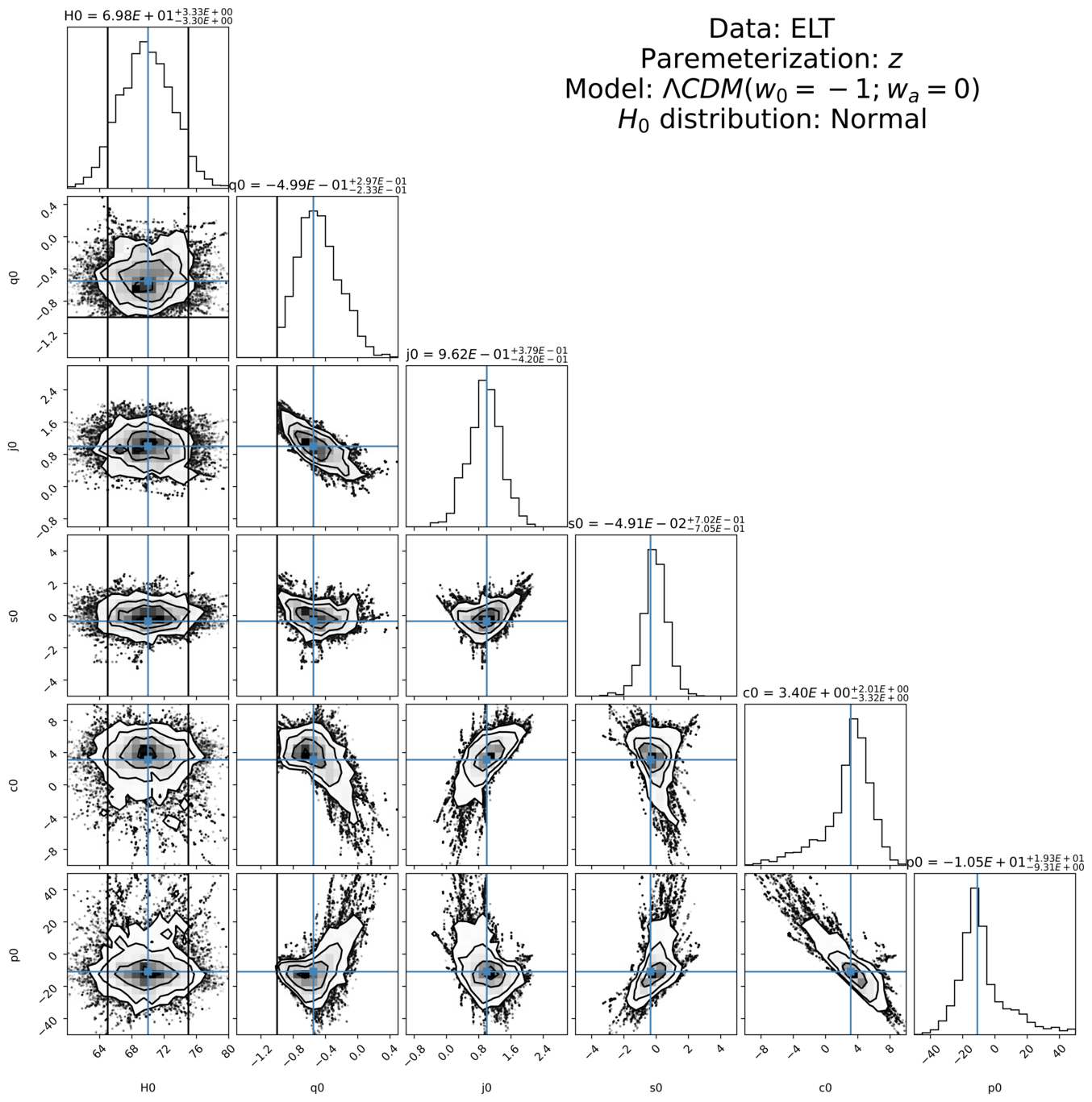}
\includegraphics[width=0.45\columnwidth,keepaspectratio]{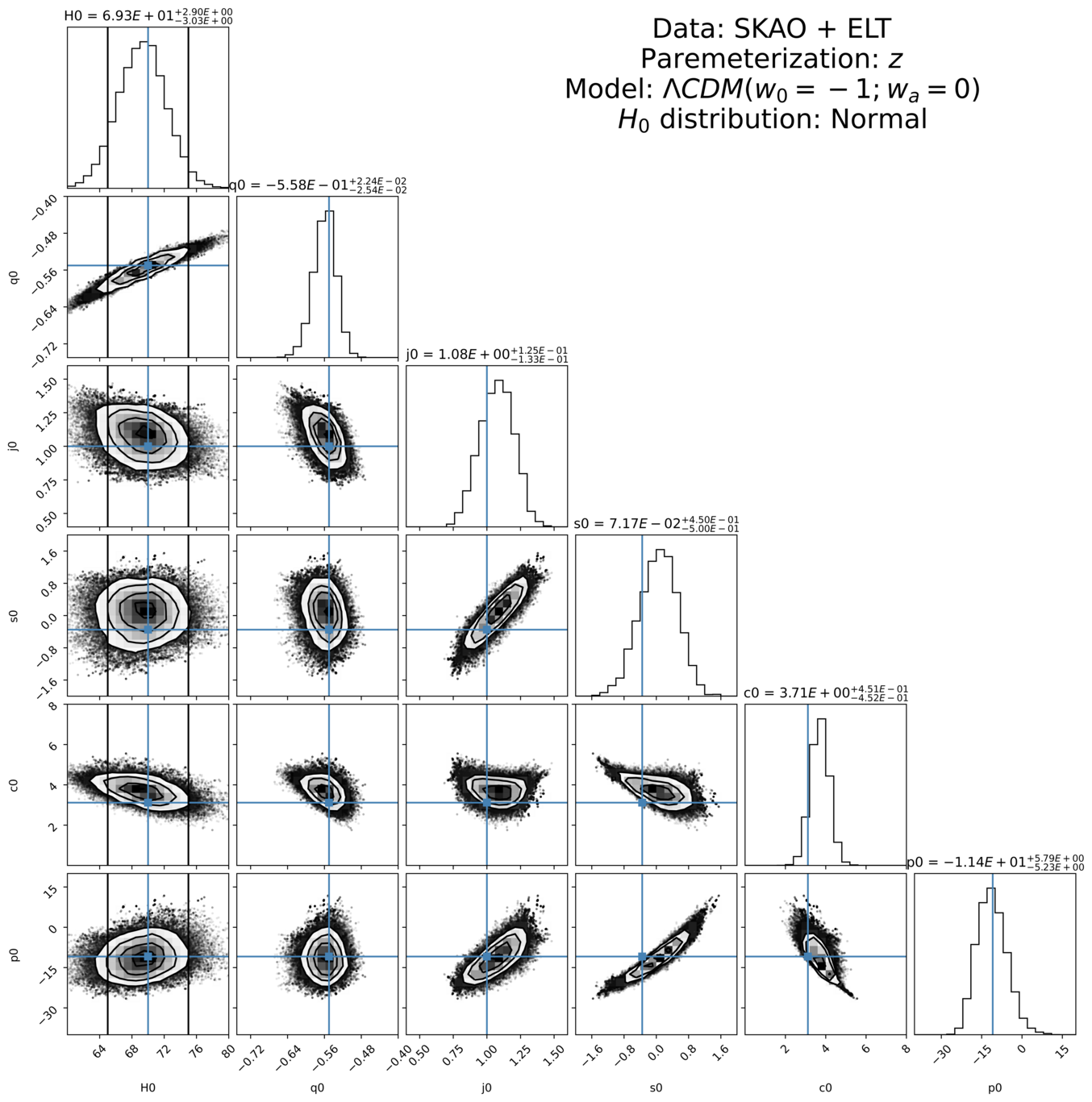}
\includegraphics[width=0.45\columnwidth,keepaspectratio]{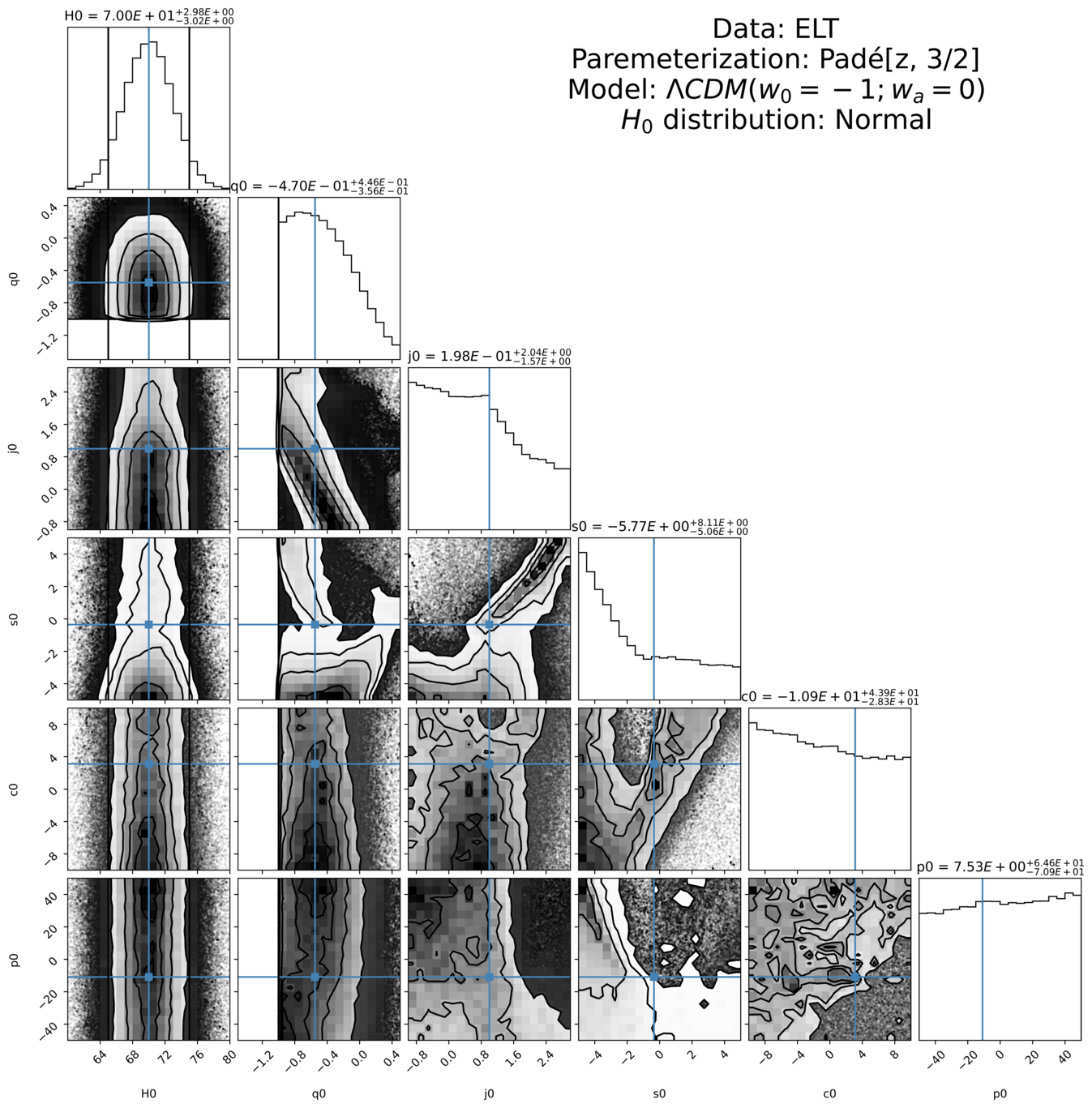}
\includegraphics[width=0.45\columnwidth,keepaspectratio]{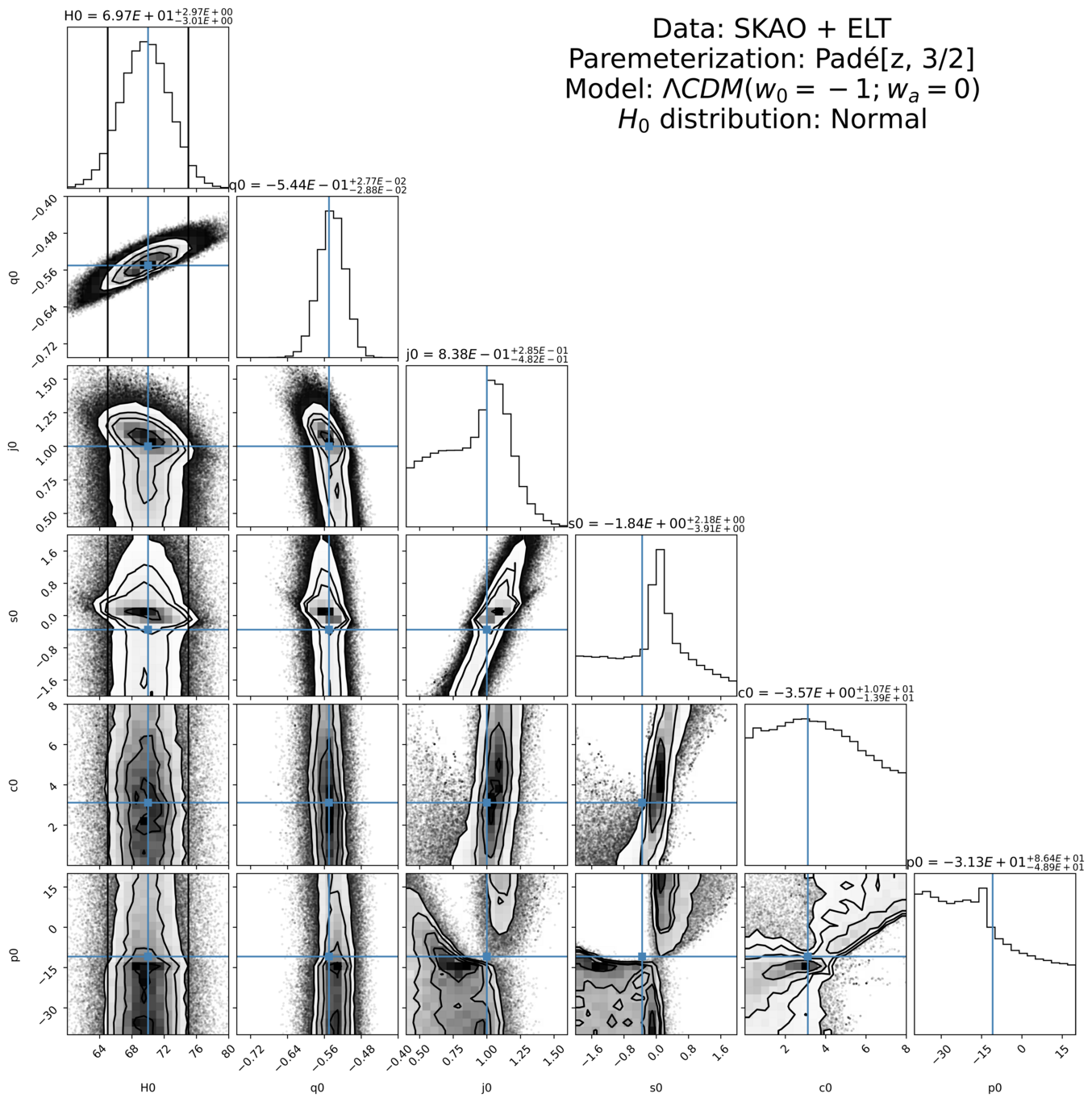}
\includegraphics[width=0.45\columnwidth,keepaspectratio]{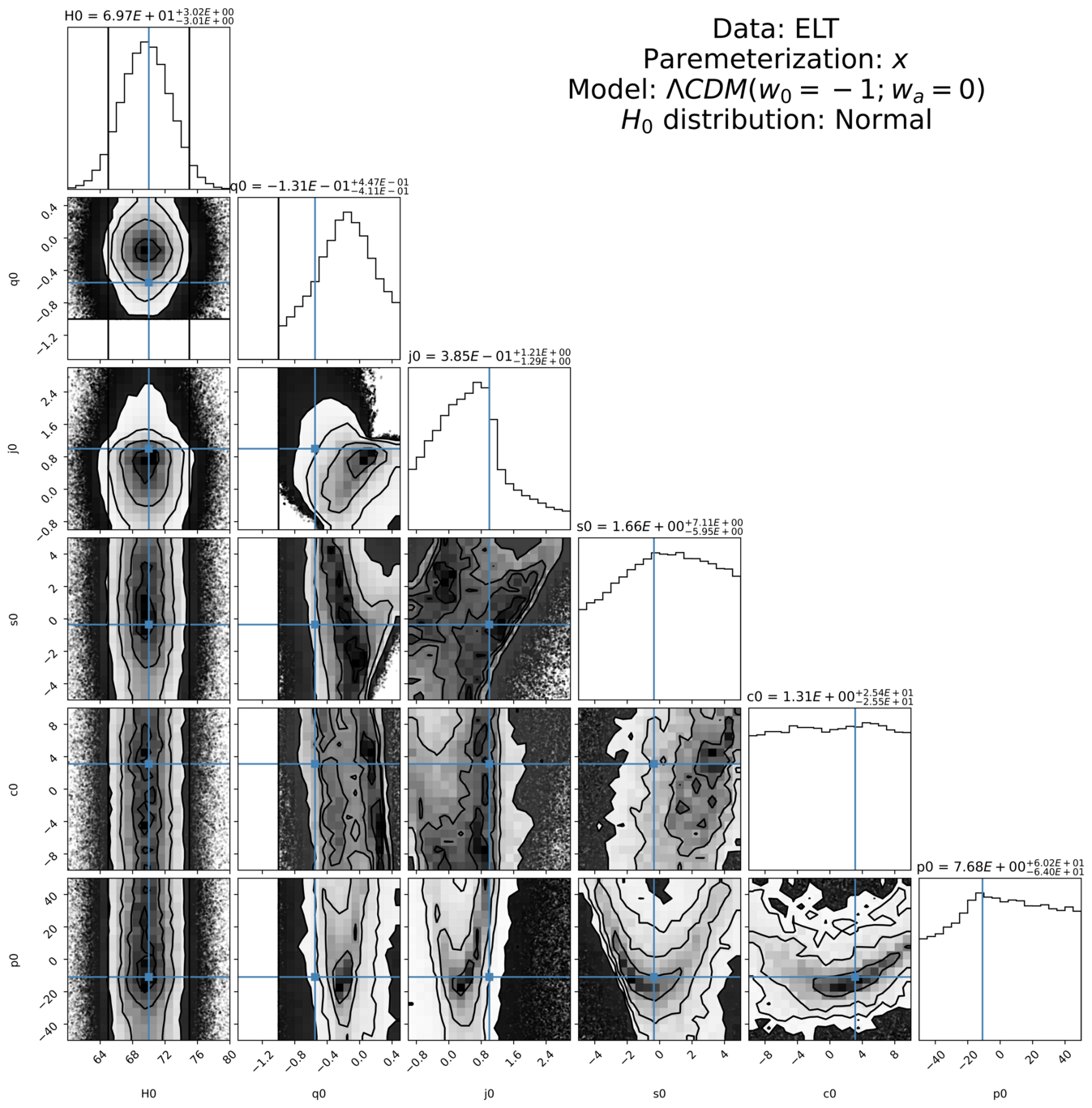}
\includegraphics[width=0.45\columnwidth,keepaspectratio]{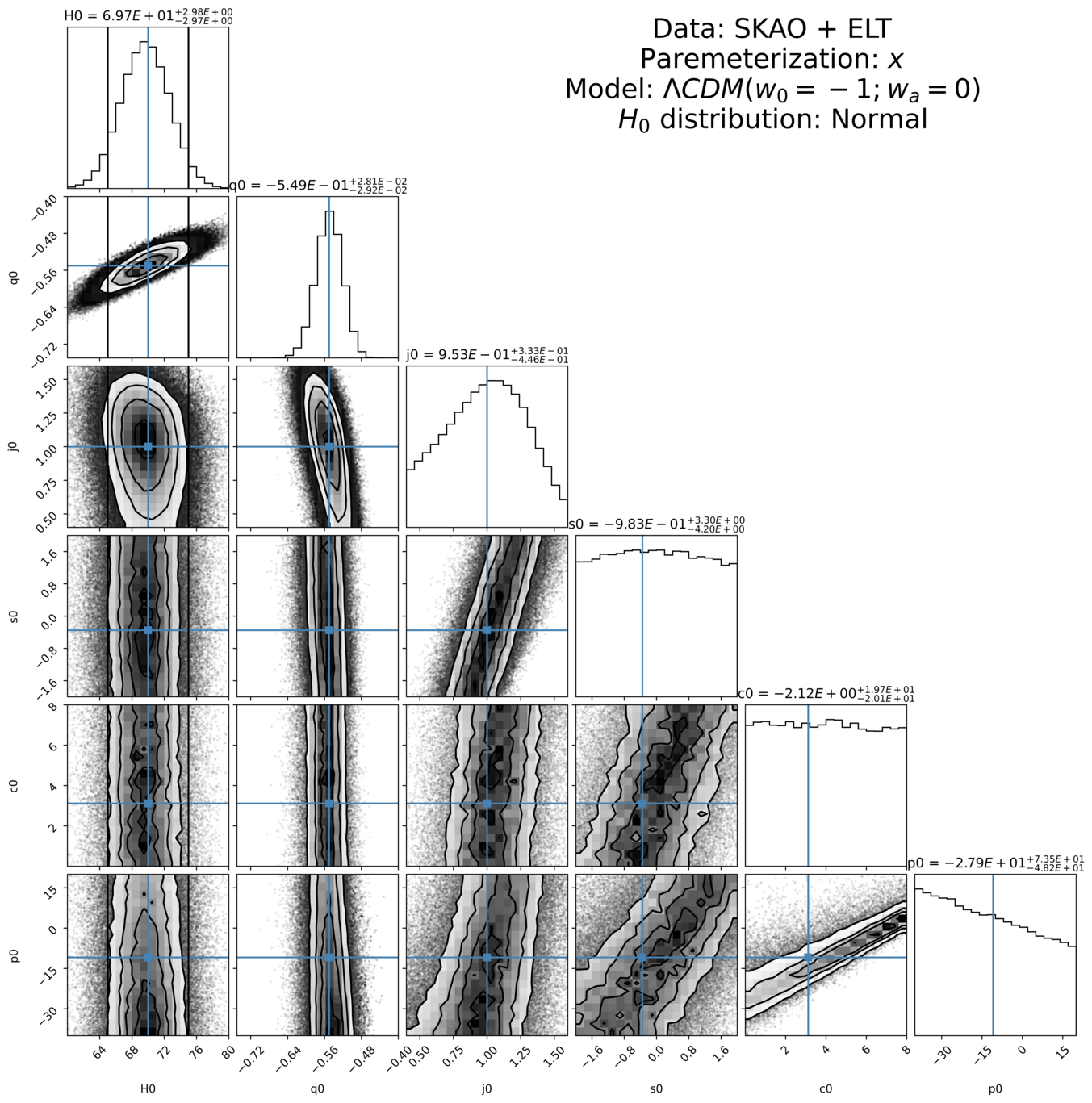}
\end{center}
\caption{Cosmographic parameters from ELT data with the $z$, Pad\'e$[z,3/2]$ and $x$ series (top, middle and bottom panels), with normal priors on $H_0$. Left panels are for the ELT alone, right ones for SKAO+ELT. The blue lines identify the true (fiducial model) values for each parameter. For $H_0$, the black lines identify the  range of values between 65 and 75 km/s/Mpc; for the other parameters, the black lines (when visible) identify the limits of our priors.}
\label{fig03}
\end{figure}

A similar analysis can now be done for the case of the ELT data. The results obtained for the ELT alone are shown in the left-side panels of Figure \ref{fig03} and in the left part of Table \ref{tab4}. As previously mentioned, we only show the results for the case of a normal prior on the Hubble constant.

\begin{table}
\centering
\caption{Constraints on the cosmographic coefficients of the three series expansions, obtained from the MCMC analysis of the ELT data set alone (left side) and the combination of SKAO and ELT data (right side). Both use a normal prior on the Hubble constant. The middle column shows the correct values (approximated to two decimal places) for the assumed fiducial model.}\label{tab4}
\begin{tabular}{ |c|c|c|c|c|c|c|  }
    \hline
     \multicolumn{3}{|c|}{Results for ELT alone} & \multicolumn{1}{|c|}{Parameter \& Fiducial value} & \multicolumn{3}{|c|}{Results for SKAO+ELT} \\
    $z$ & $P[z,3/2]$ & $x$ & { } & $z$ & $P[z,3/2]$ & $x$ \\
    \hline
    $69.80^{+3.33}_{-3.30}$ & $69.99^{+2.98}_{-3.02}$ & $69.68^{+3.02}_{-3.01}$ & $H_0=70$ & $69.30^{+2.90}_{-3.03}$ & $69.67^{+2.97}_{-3.01}$ & $69.66^{+2.98}_{-2.97}$\\
    $-0.50^{+0.30}_{-0.23}$ & $-0.47^{+0.45}_{-0.36}$ & $-0.13^{+0.45}_{-0.41}$ & $q_0=-0.55$ & $-0.56^{+0.02}_{-0.03}$ & $-0.54^{+0.03}_{-0.03}$ & $-0.55^{+0.03}_{-0.03}$\\
    $0.96^{+0.38}_{-0.42}$ & $0.20^{+2.04}_{-1.57}$ & $0.38^{+1.21}_{-1.29}$ & $j_0=1$ & $1.08^{+0.12}_{-0.13}$ & $0.84^{+0.29}_{-0.48}$ & $0.95^{+0.33}_{-0.45}$ \\
    $-0.05^{+0.70}_{-0.71}$ & $-5.77^{+8.11}_{-5.06}$ & $1.66^{+7.11}_{-5.95}$ & $s_0=-0.35$ & $0.07^{+0.45}_{-0.50}$ & $-1.84^{+2.18}_{-3.91}$ & $-0.98^{+3.30}_{-4.20}$\\
    $3.40^{+2.01}_{-3.32}$ & $-10.94^{+43.92}_{-28.25}$ & $1.31^{+25.35}_{-25.49}$ & $c_0=3.12$ & $3.71^{+0.45}_{-0.45}$ & $-3.57^{+10.66}_{-13.87}$ & $-2.12^{+19.67}_{-20.10}$\\
    $-10.51^{+19.32}_{-9.31}$ & $7.53^{+64.59}_{-70.89}$ & $7.68^{+60.20}_{-63.95}$ & $p_0=-10.89$ & $-11.42^{+5.79}_{-5.23}$ & $-31.34^{+86.44}_{-48.86}$ & $-27.95^{+73.50}_{-48.18}$\\
    \hline
\end{tabular}
\end{table}

Qualitatively, we may approach this case by considering that we have 5 measurements of the redshift drift, with which we can determine the 5 free parameters (the sixth, $H_0$, is effectively determined by the prior), and therefore this should always be possible. For the $z$ series, this is indeed the case, and (again qualitatively) each term in the series determines one parameter: the linear term determines $q_0$, the quadratic term---which depends on $(j_0-q_0^2)$---then determines $j_0$, and so on. It follows that all one-dimensional posteriors are sharply peaked around the fiducial values, and without biases, as can be seen in the top left plot of Fig. \ref{fig03}. However, for the other series the situation is not as simple. This highlights one of their shortcomings, which is the more complicated dependence of the series on the parameters. For example, for the $x$ series the quadratic term depends on $(j_0-q_0^2+q_0)$, and this leads to more degeneracies and actually results in a biased $q_0$ for this series, which is clearly visible in the bottom left plot. The situation is even worse for the Pad\'e case, which, although without biases, leads to significantly larger uncertainties, together with extremely messy likelihoods.

It is noteworthy that all cosmographic expansions have some difficulty in predicting values for the deceleration parameter $q_0$, although the reason for this can be easily understood. The deceleration parameter is the only one that appears on its own in one of the terms in the series---the linear one (ignoring the fact that $H_0$ is an overall multiplicative factor). It follows that if there is low-redshift data, a reasonable determination of $q_0$ is always possible, and will be mainly due to this term \citep{Neben}. On the other hand, if one only has high-redshift data and must necessarily include several more terms in the series (which all depend on different combinations of the parameters), then $q_0$ will, so to speak, have to compete with the other parameters. Thus, instead of having a clear hierarchy of uncertainties (with the lower order parameters well constrained and the higher order ones much less so), the differences in uncertainties of the various terms are, proportionally, much smaller. This is another reason why, for this kind of observational data set, the comparatively simpler parameter space (in terms of degeneracies between the relevant parameters) of the $z$ series is advantageous.

We can also consider the combination of the SKAO and ELT data sets. These results, again obtained with a normal prior on the Hubble constant, are shown in the right-side panels of Figure \ref{fig03} and in the right part of Table \ref{tab4}. In this case, as one might expect, there are some gains in sensitivity from the combination of the two data sets. These gains are especially large for the $z$ series, which yields significantly tighter constraints than the other two series---indeed, even $p_0$ is well constrained in this case. However, this comes at one cost; there is a bias in $c_0$ (in the sense defined above), while the other series are, in this sense, unbiased. For the other two series, the gains with respect to the case of SKAO data only are more noticeable in the intermediate parameters in the series---in other words, the jerk and the snap.

\subsection{Interlude: An ideal experiment}

While in the rest of the work we have aimed to use realistic simulated data, which is representative of what the SKAO and the ELT will achieve (at least to the extent that this can be assessed at the moment), in this section we allow ourselves to be entirely speculative, and consider a hypothetical scenario where we would have 100 SKAO measurements and 100 ELT measurements, evenly distributed in the same redshift ranges as specified in Sect. \ref{datasets} and also using the same prescription for the redshift-dependent measurement uncertainties described therein. While this would be prohibitively expensive in terms of telescope time in both facilities, the goal of the exercise is to assess the possible limitations of the cosmographic approach to the redshift drift. In other words, this will help in identifying which of these are intrinsically due to the method itself (and are therefore unavoidable bottlenecks) and which are simply due to the limited quality and/or quantity of the data.

\begin{figure}
\begin{center}
\includegraphics[width=0.32\columnwidth,keepaspectratio]{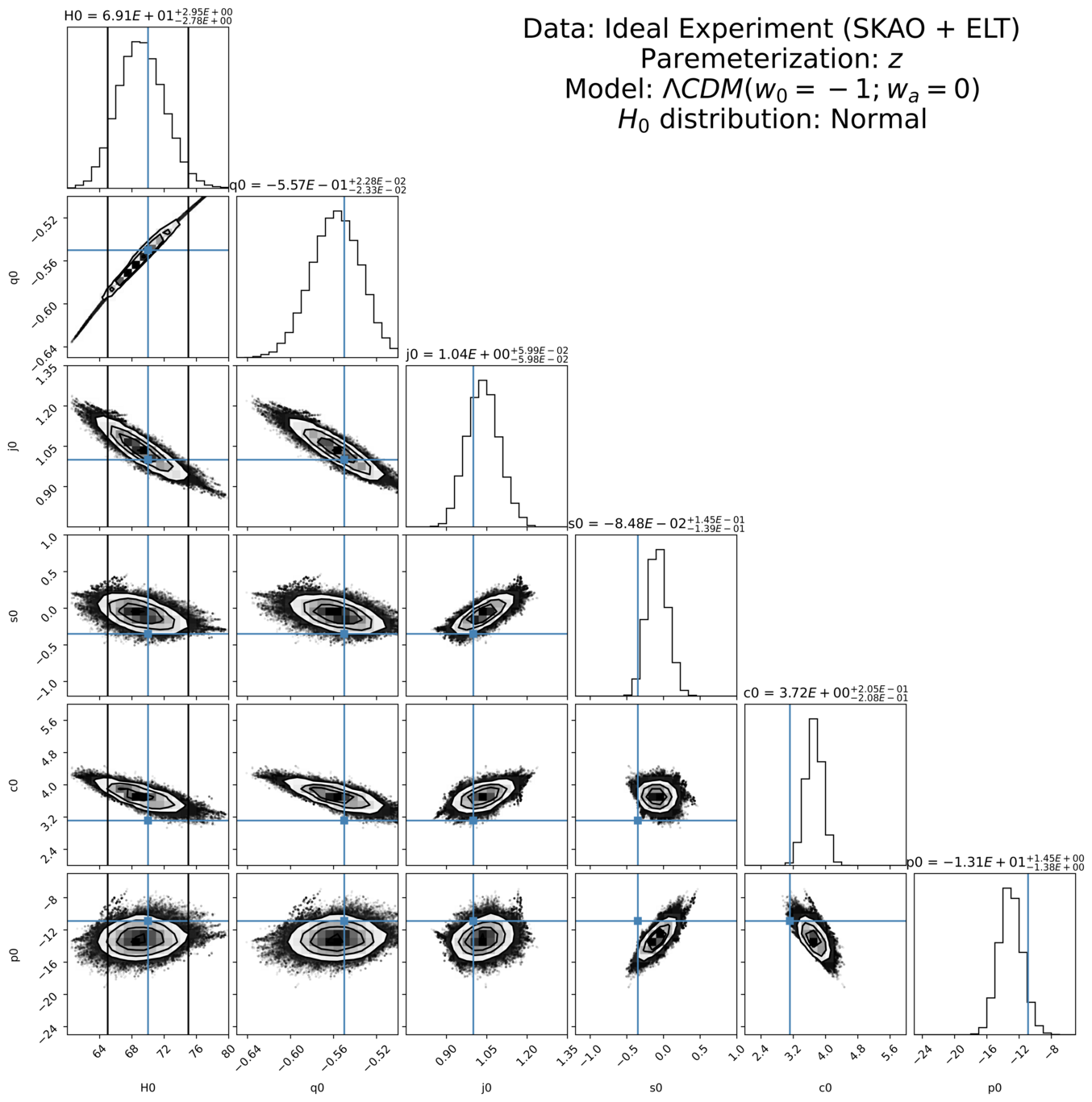}
\includegraphics[width=0.32\columnwidth,keepaspectratio]{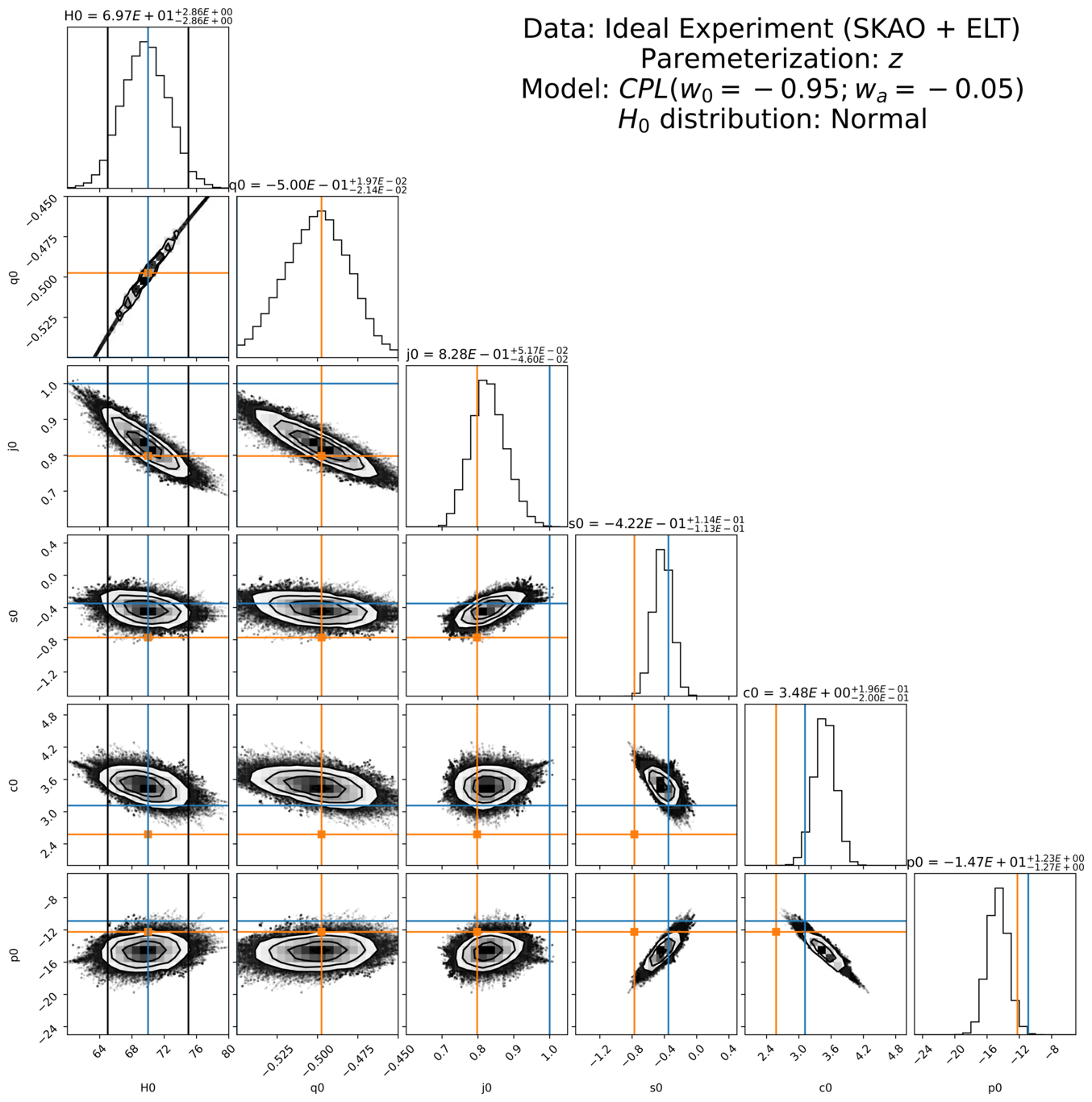}
\includegraphics[width=0.32\columnwidth,keepaspectratio]{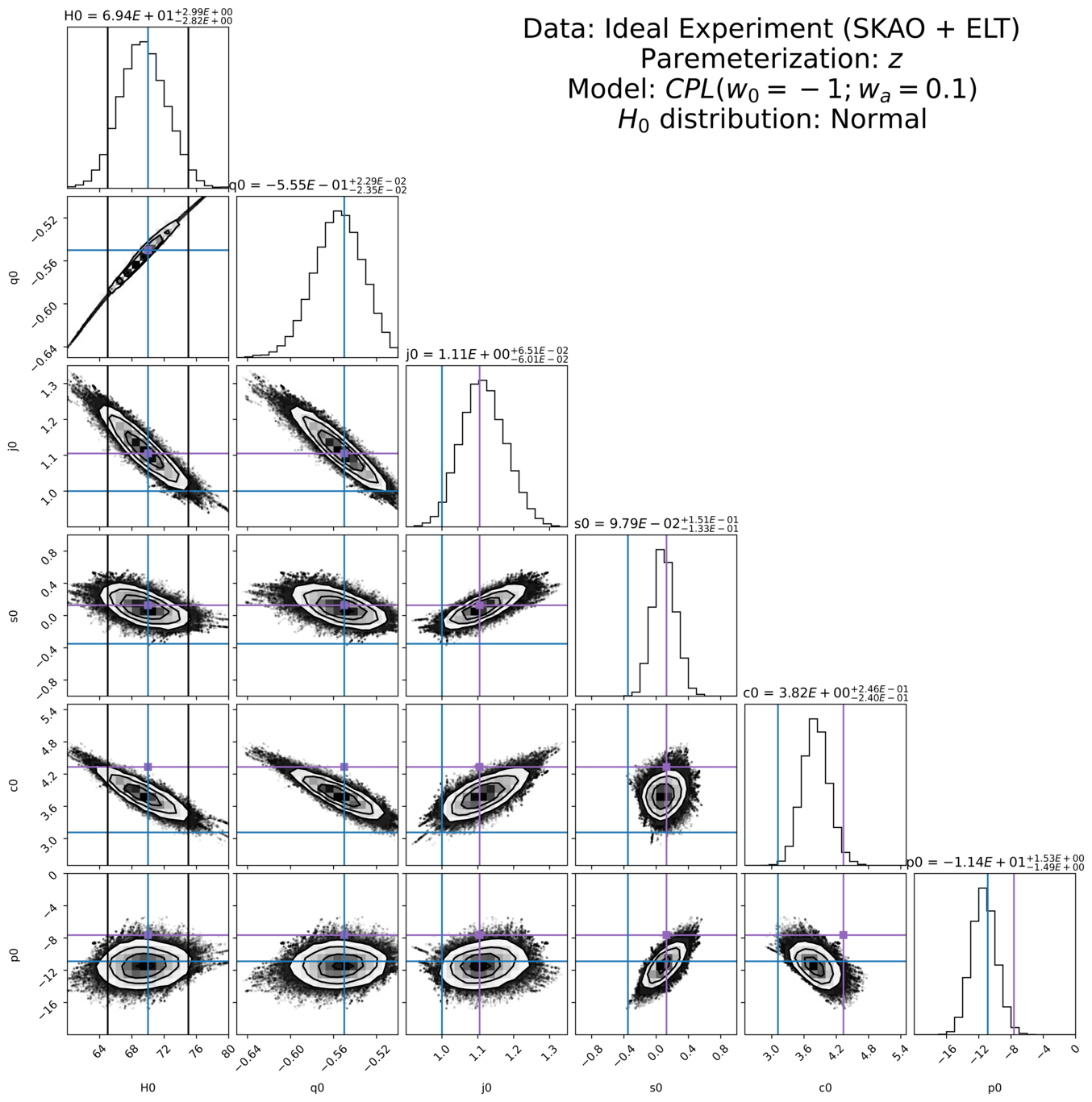}
\includegraphics[width=0.32\columnwidth,keepaspectratio]{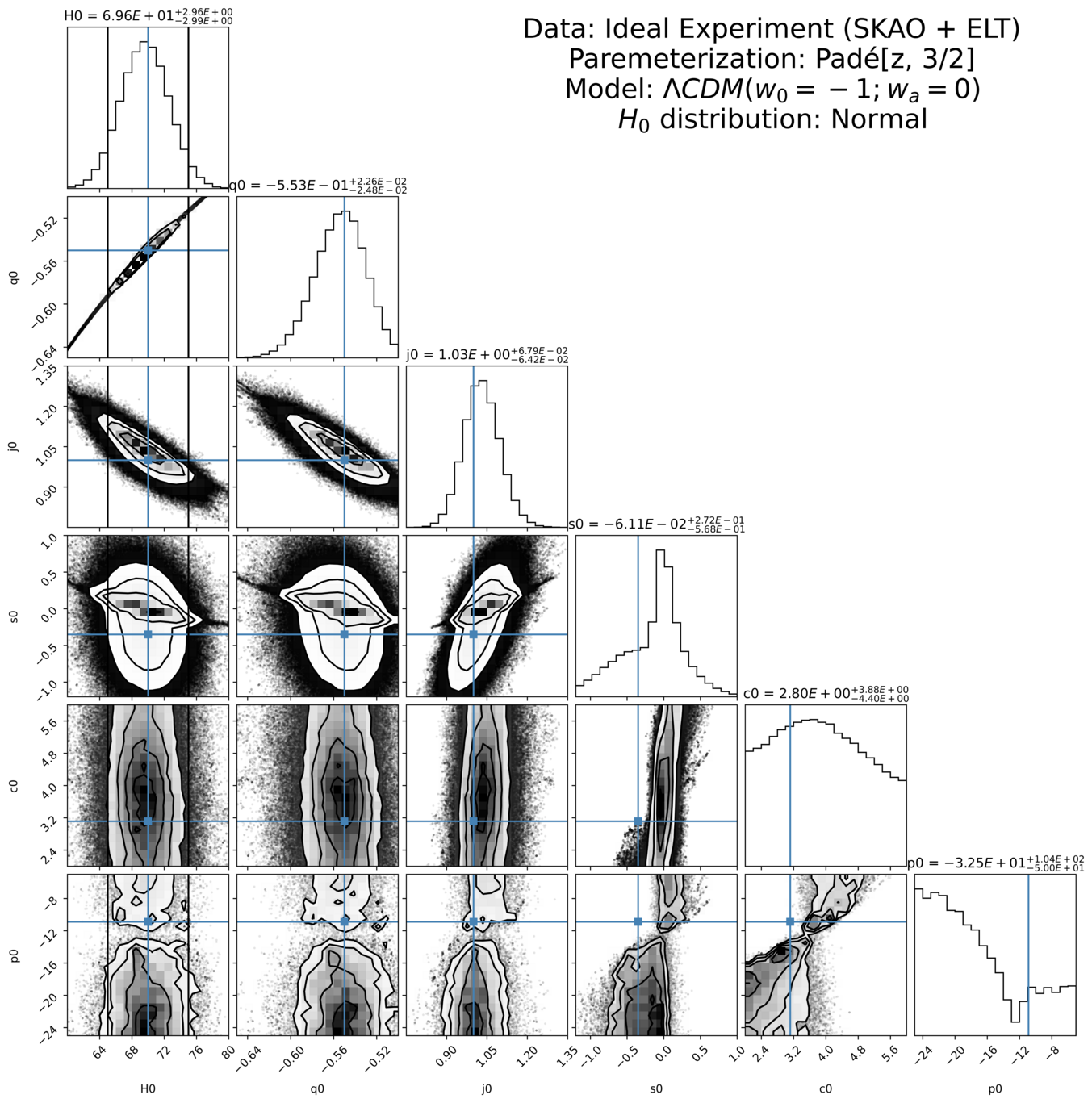}
\includegraphics[width=0.32\columnwidth,keepaspectratio]{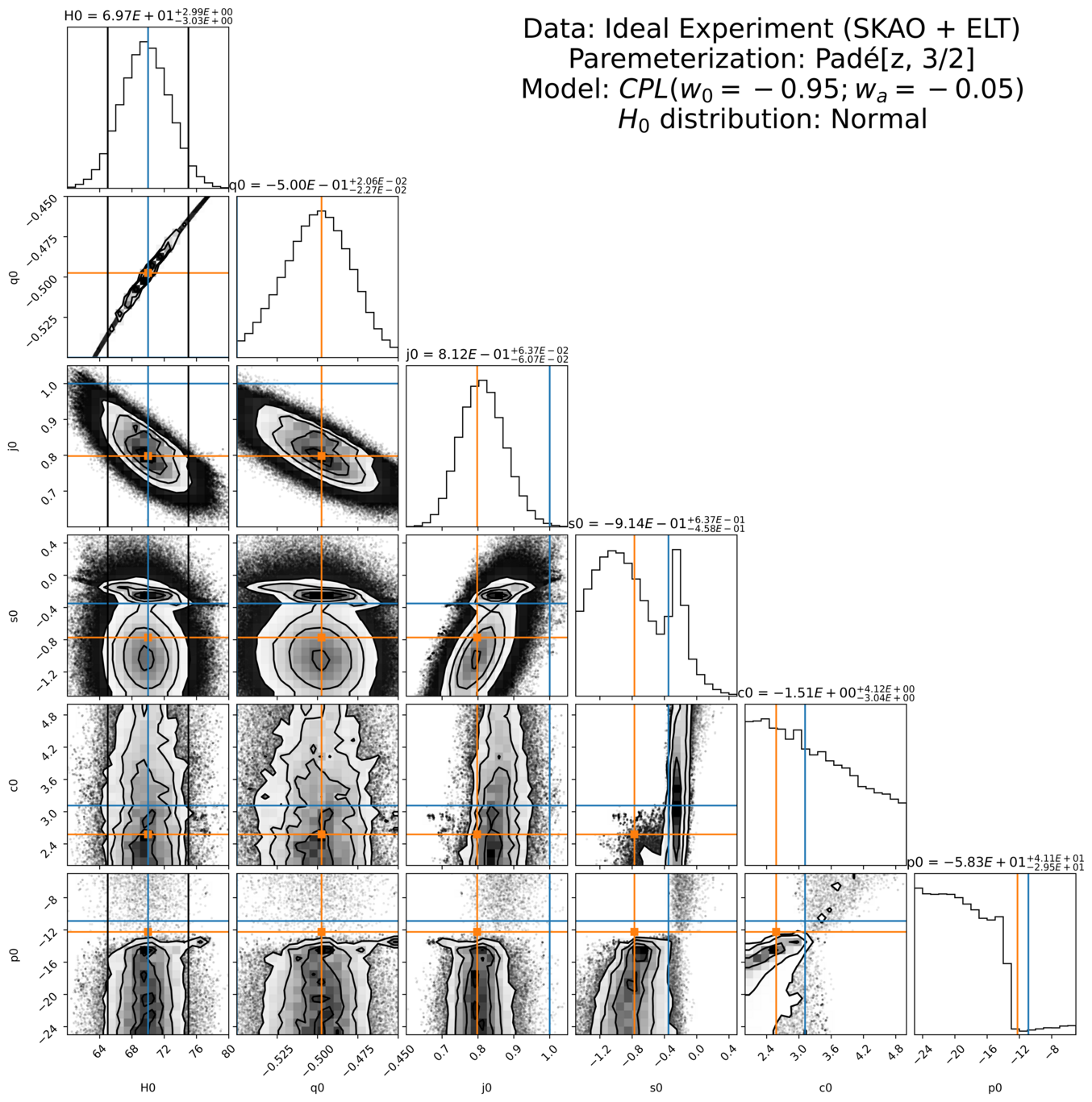}
\includegraphics[width=0.32\columnwidth,keepaspectratio]{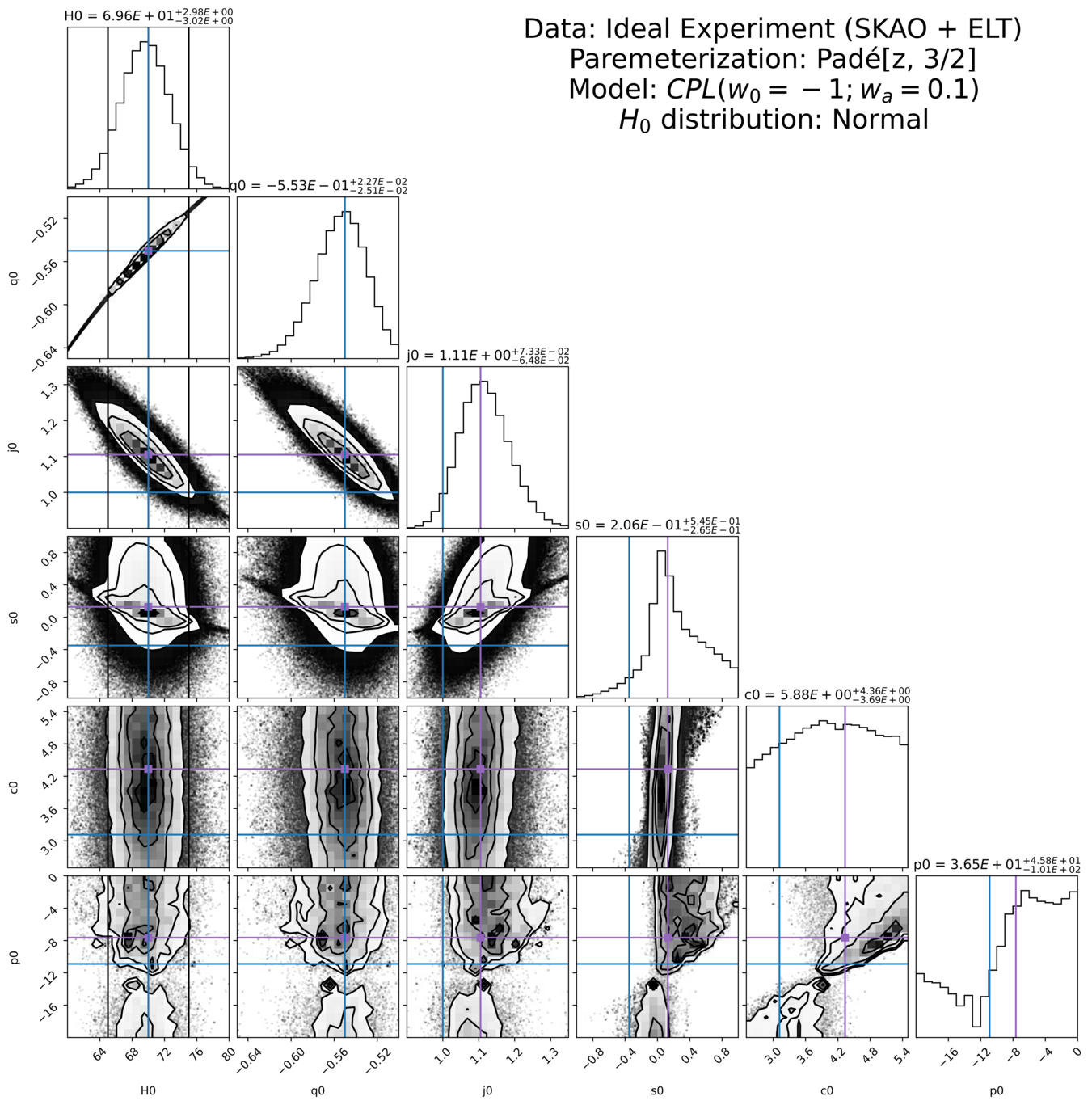}
\includegraphics[width=0.32\columnwidth,keepaspectratio]{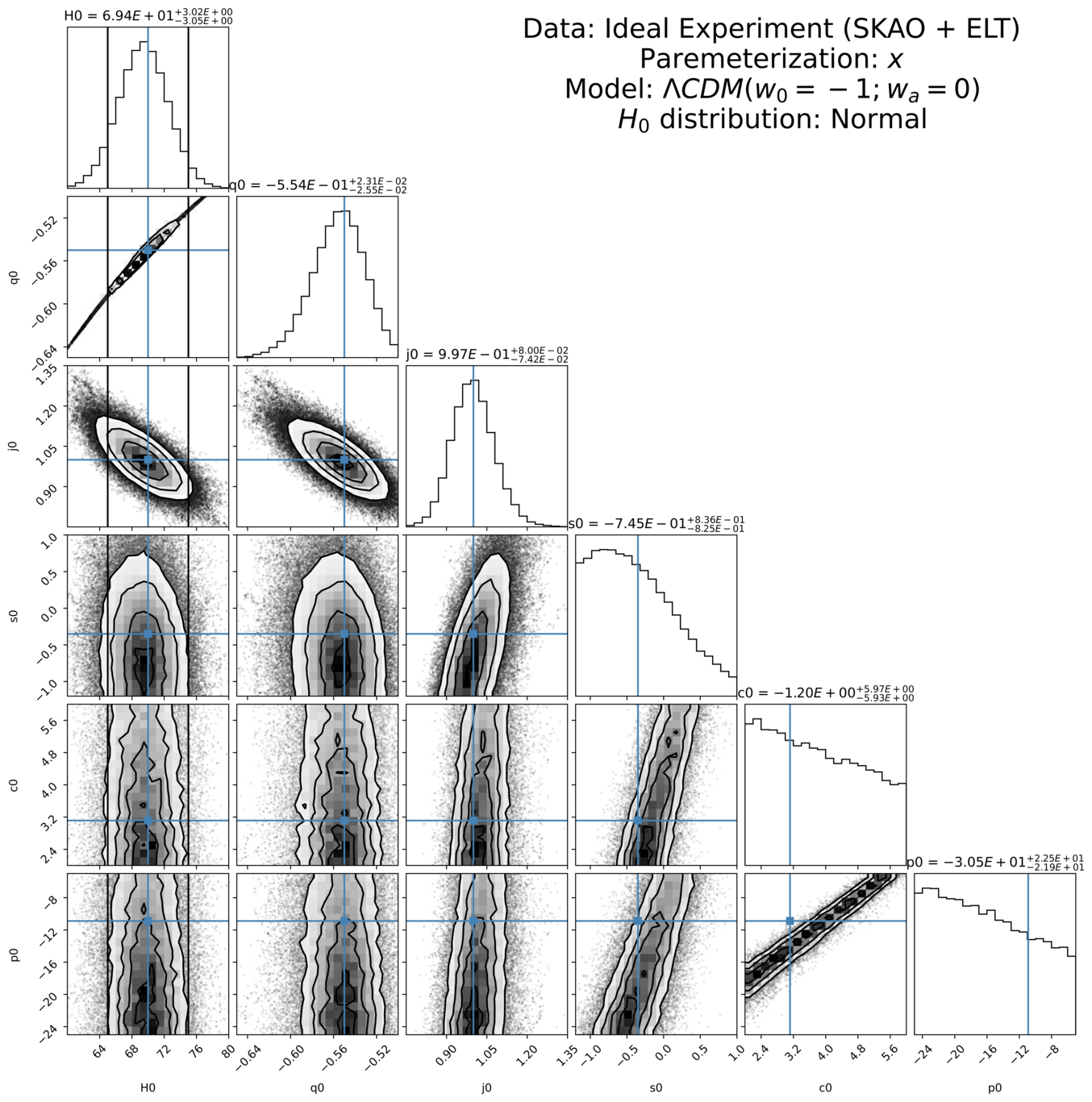}
\includegraphics[width=0.32\columnwidth,keepaspectratio]{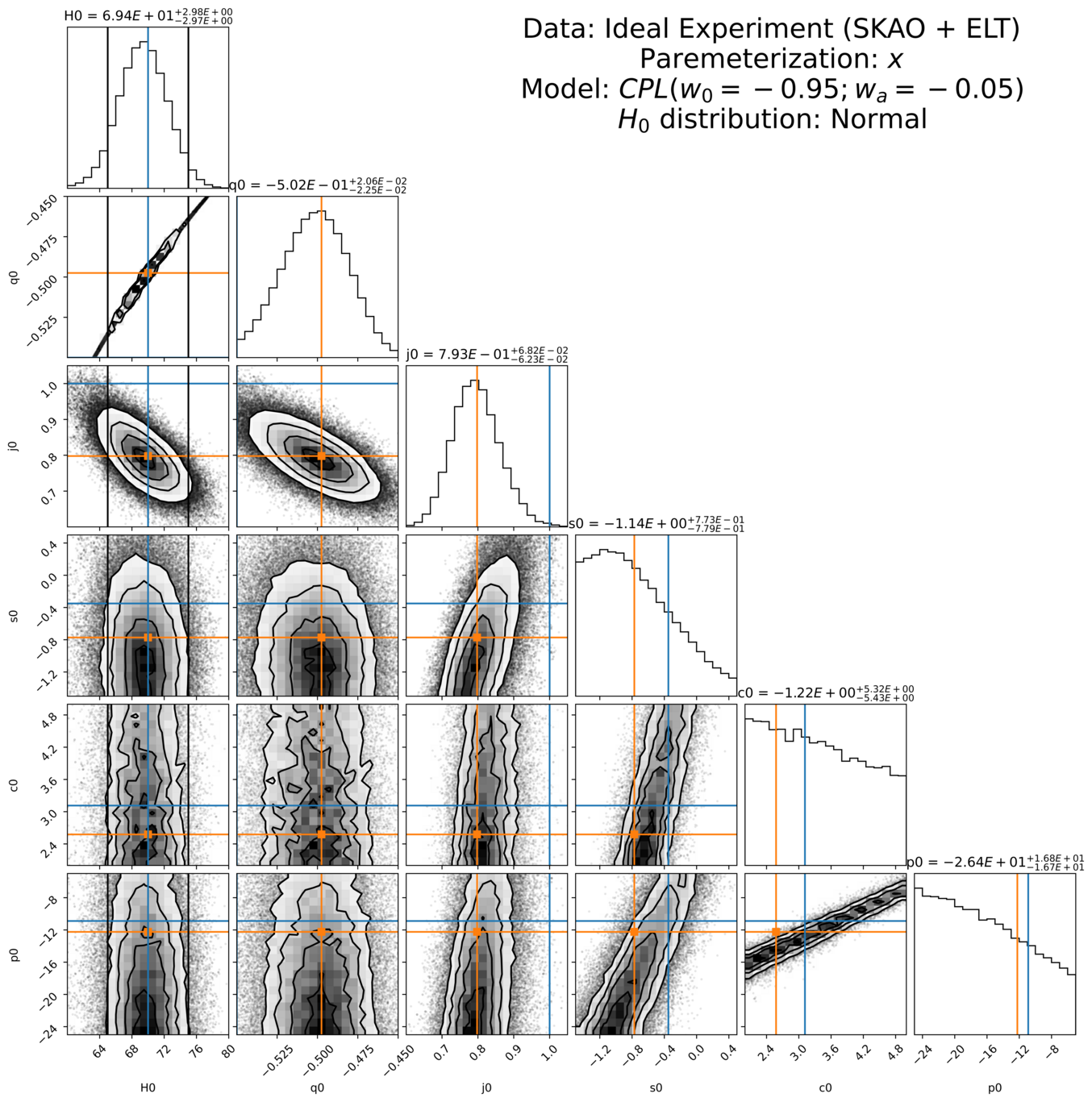}
\includegraphics[width=0.32\columnwidth,keepaspectratio]{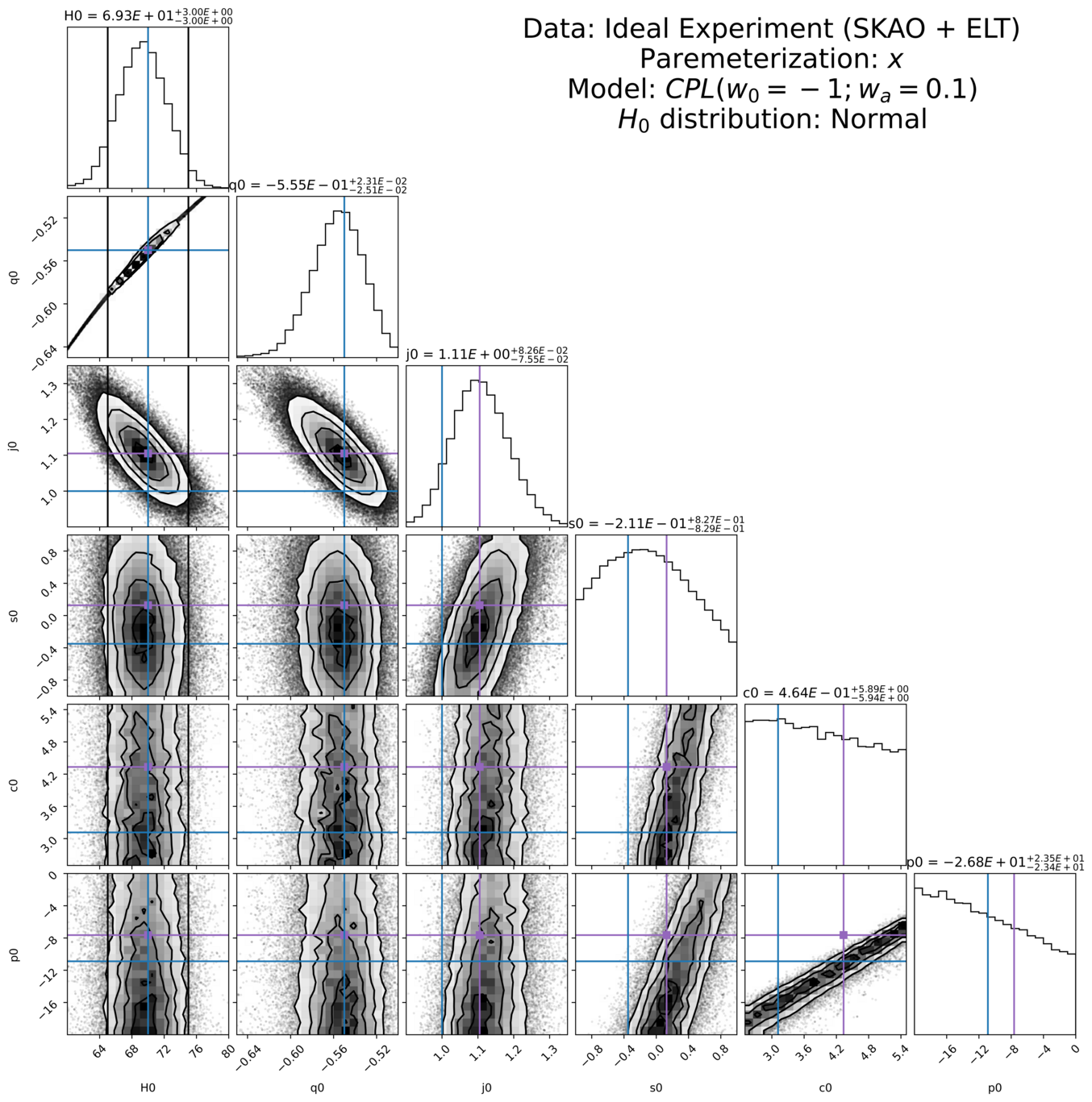}
\end{center}
\caption{Cosmographic parameters from ideal data with the $z$, Pad\'e$[z,3/2]$ and $x$ series (top, middle and bottom panels), with normal priors on $H_0$. Left, middle and right panels are for the flat $\Lambda$CDM and the CPL $(w_0=-0.95,w_a=-0.05)$ and $(w_0=-1.0, w_a=0.1)$ fiducial models respectively. The blue, orange. and purple lines identify, respectively, the true (fiducial model) values for the parameters in each of the three cases. For $H_0$, the black lines identify the  range of values between 65 and 75 km/s/Mpc; for the other parameters, the black lines (when visible) identify the limits of our priors.}
\label{fig04}
\end{figure}
\begin{table}
\centering
\caption{Constraints on the cosmographic coefficients of the three series expansions, obtained from the MCMC analysis of the ideal data sets. The top, middle and bottom parts of the table are for the flat $\Lambda$CDM and the CPL $(w_0=-0.95,w_a=-0.05)$ and $(w_0=-1.0, w_a=0.1)$ fiducial models respectively. A normal prior on the Hubble constant is used throughout, and the correct values of the coefficients for the assumed fiducial model are approximated to two decimal places.}\label{tab5}
\begin{tabular}{ |c|c|c|c|c|  }
    \hline
    Fiducial Model & Fiducial parameter & $z$ & $P[z,3/2]$ & $x$ \\
    \hline
    {} & $H_0=70$ & $69.12^{+2.95}_{-2.78}$ & $69.62^{+2.96}_{-2.99}$ & $69.42^{+3.02}_{-3.05}$ \\
    {$\Lambda$CDM} & $q_0=-0.55$ & $-0.56^{+0.02}_{-0.02}$ & $-0.55^{+0.02}_{-0.02}$ & $-0.55^{+0.02}_{-0.03}$ \\
    {} & $j_0=1$ & $1.04^{+0.06}_{-0.06}$ & $1.03^{+0.07}_{-0.06}$ & $1.00^{+0.08}_{-0.07}$ \\
    {$w_0=-1$} & $s_0=-0.35$ & $-0.08^{+0.14}_{-0.14}$ & $-0.06^{+0.27}_{-0.57}$ & $-0.75^{+0.84}_{-0.82}$ \\
    {$w_a=0$} & $c_0=3.12$ & $3.72^{+0.20}_{-0.21}$ & $2.80^{+3.88}_{-4.40}$ & $-1.20^{+5.97}_{-5.93}$ \\
    {} & $p_0=-10.89$ & $-13.07^{+1.45}_{-1.38}$ & $-32.54^{+103.81}_{-49.97}$ & $-30.51^{+22.52}_{-21.86}$ \\
    \hline
    {} & $H_0=70$ & $69.66^{+2.86}_{-2.86}$ & $69.67^{+2.99}_{-3.03}$ & $69.40^{+2.98}_{-2.97}$\\
    {CPL} & $q_0=-0.50$ & $-0.50^{+0.02}_{-0.02}$ & $-0.50^{+0.02}_{-0.02}$ & $-0.50^{+0.02}_{-0.02}$\\
    {} & $j_0=0.80$ & $0.83^{+0.05}_{-0.05}$ & $0.81^{+0.06}_{-0.06}$ & $0.79^{+0.07}_{-0.06}$\\
    {$w_0=-0.95$}& $s_0=-0.77$ & $-0.42^{+0.11}_{-0.11}$ & $-0.91^{+0.64}_{-0.46}$ & $-1.14^{+0.77}_{-0.78}$\\
    {$w_a=-0.05$} & $c_0=2.58$ & $3.48^{+0.20}_{-0.20}$ & $-1.51^{+4.12}_{-3.04}$ & $-1.22^{+5.32}_{-5.43}$\\
    {} & $p_0=-12.24$ & $-14.65^{+1.23}_{-1.27}$ & $-58.31^{+41.09}_{-29.49}$ & $-26.44^{+16.76}_{-16.73}$\\
    \hline
    {} & $H_0=70$ & $69.39^{+2.99}_{-2.82}$ & $69.62^{+2.98}_{-3.02}$ & $69.32^{+3.00}_{-3.00}$\\
    {CPL} & $q_0=-0.55$ & $-0.55^{+0.02}_{-0.02}$ & $-0.55^{+0.02}_{-0.03}$ & $-0.56^{+0.02}_{-0.03}$\\
    {} & $j_0=1.11$ & $1.11^{+0.07}_{-0.06}$ & $1.11^{+0.07}_{-0.06}$ & $1.11^{+0.08}_{-0.08}$\\
    {$w_0=-1$} & $s_0=0.13$ & $0.10^{+0.15}_{-0.13}$ & $0.21^{+0.54}_{-0.27}$ & $-0.21^{+0.83}_{-0.83}$\\
    {$w_a=0.1$} & $c_0=4.33$ & $3.82^{+0.25}_{-0.24}$ & $5.88^{+4.36}_{-3.69}$ & $0.46^{+5.89}_{-5.94}$\\
    {} & $p_0=-7.64$ & $-11.36^{+1.53}_{-1.49}$ & $36.47^{+45.83}_{-101.01}$ & $-26.80^{+23.46}_{-23.42}$\\    
    \hline
    \end{tabular}
\end{table}

The results for this case are shown in the left side panels of Figure \ref{fig04} and on top third of Table \ref{tab5}. Overall, we find a situation similar to the one discussed at the end of the previous subsection. The $z$ series nominally leads to the tighter constraints on the series coefficients (and for the three higher-order parameters the uncertainties are much smaller than those in the other series), but this is at the cost of biases---in this case, the inferred values of $s_0$, $c_0$ and $p_0$ are all biased. The $x$ series, despite yielding larger error bars, is unbiased and numerically stable, while the Pad\'e$[z,3/2]$ still displays the aforementioned numerical instabilities.

Finally, as a simple way to quantify the model discriminating capabilities of cosmography, we consider a comparison between the flat $\Lambda$CDM fiducial model which we have taken as our baseline, with $\Omega_m0=0.3$ and $H_0=70$ km/s/Mpc, with two other flat CPL models with the same matter density and Hubble constant but different dark energy parameters. The first of these has $(w_0=-0.95,w_a=-0.05)$, and was also used in Sect. \ref{fiducial}, while the second has $(w_0=-1.0, w_a=0.1)$. Both of these models will have very similar expansion histories to the $\Lambda$CDM one, and the same Hubble constant, but the difference between their other cosmographic parameters is comparatively more noticeable. Specifically, while the $\Lambda$CDM model has $(q_0=-0.55,j_0=1)$, the first of the CPL models has $(q_0\sim-0.50,j_0\sim0.80)$ while the second has $(q_0=-0.55,j_0\sim1.11)$. In other words, the first of these differs from the $\Lambda$CDM fiducial at the level of the deceleration parameter (and also at the level of the jerk), while for the second the difference only appears at a higher order in the cosmographic series, at the level of the jerk parameter.

The results for this case are shown in the middle and right side panels of Figure \ref{fig04} and on the middle and bottom thirds of Table \ref{tab5}. All the cosmographic series studied consistently show a statistical preference for the CPL model, demonstrating (at least in principle) the capacity of cosmography to distinguish between models. Overall these cases confirm the previous analysis (which is unsurprising, since both of these CPL models are fairly close, in parameter space, to the $\Lambda$CDM one). In this case the only slight difference with respect to the $\Lambda$CDM model is that for the $z$ series the snap parameter is biased for the first of the CPL models but not for the second one; the crackle and pop are still biased in both cases. Moreover, in the first CPL model $p_0$ is also biased in the Pad\'e$[z,3/2]$ case. One other lesson is that the vertical nuisance asymptotes are indeed a bottleneck of the Pad\'e approximants, even with this ideal data.

\section{Conclusions}
\label{concl}

\begin{figure}
\begin{center}
\includegraphics[width=0.45\columnwidth,keepaspectratio]{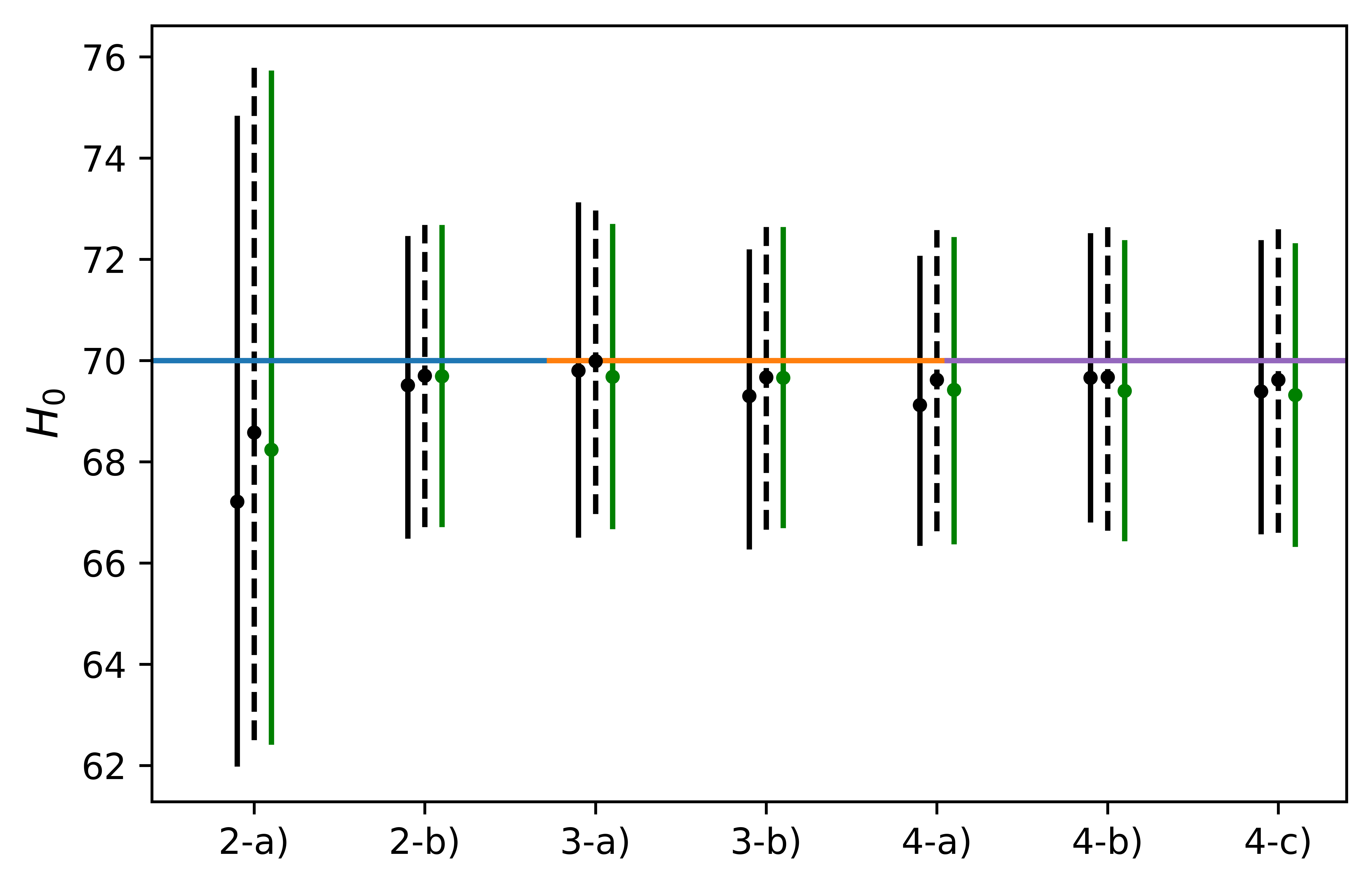}
\includegraphics[width=0.45\columnwidth,keepaspectratio]{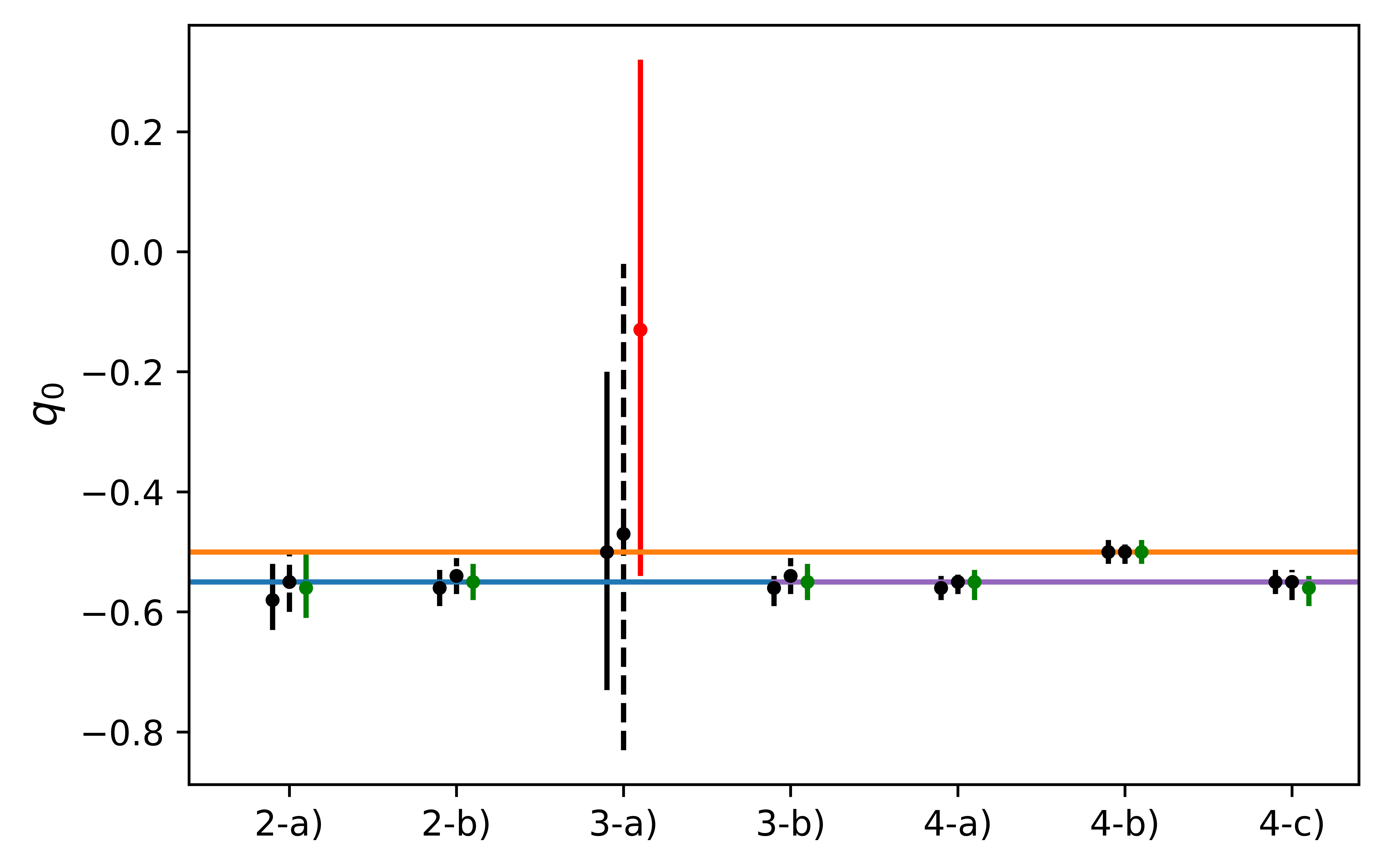}
\includegraphics[width=0.45\columnwidth,keepaspectratio]{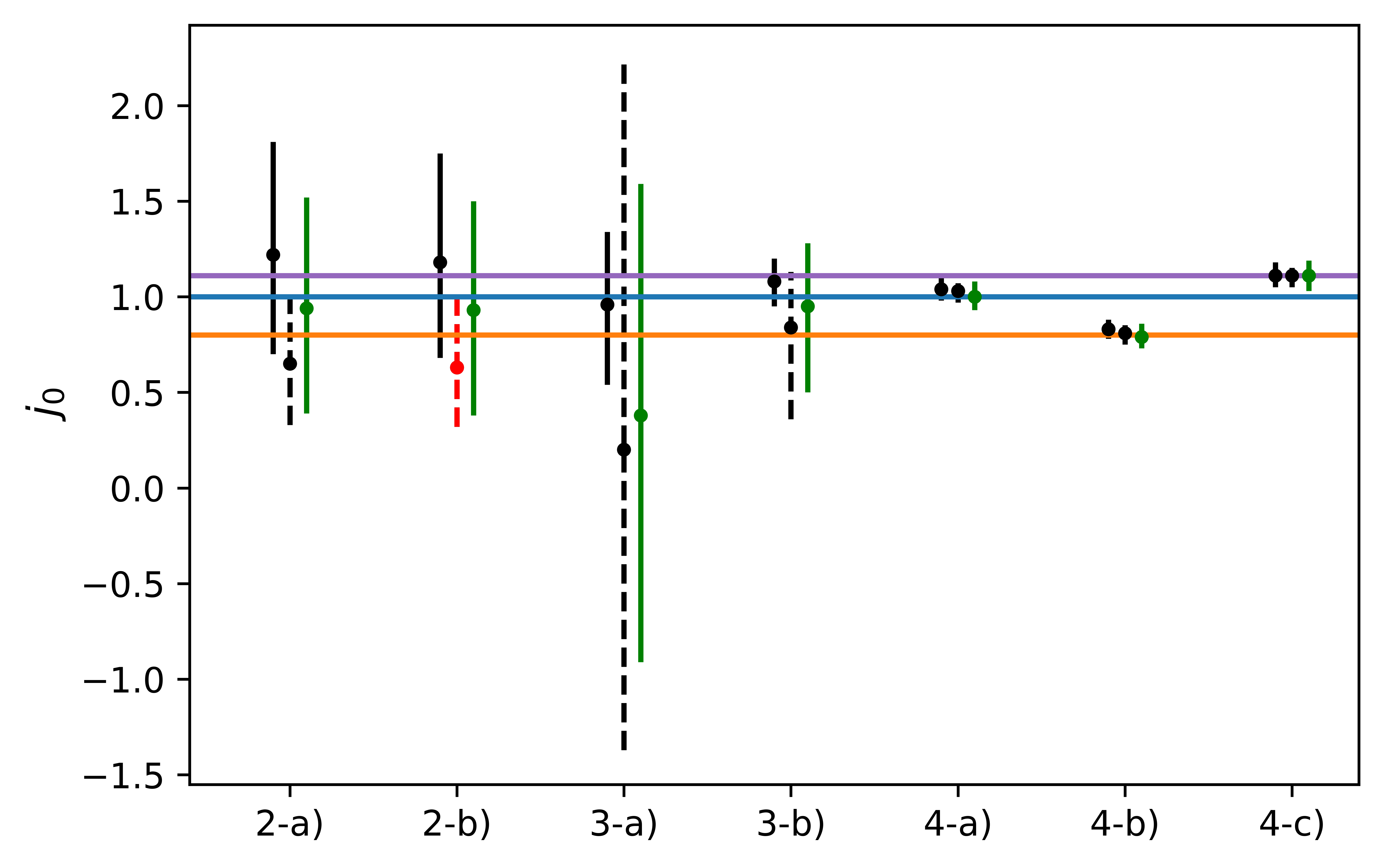}
\includegraphics[width=0.45\columnwidth,keepaspectratio]{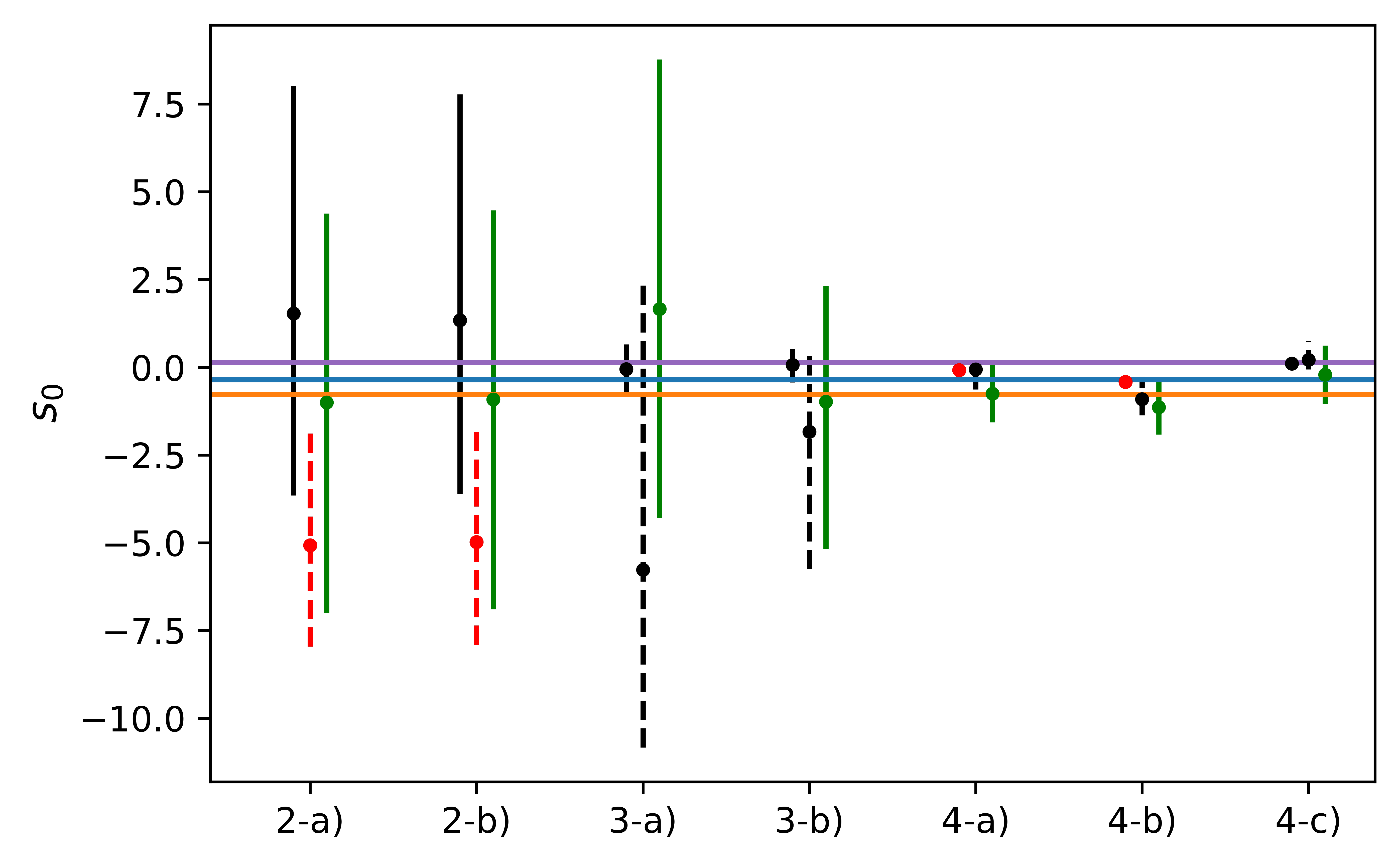}
\includegraphics[width=0.45\columnwidth,keepaspectratio]{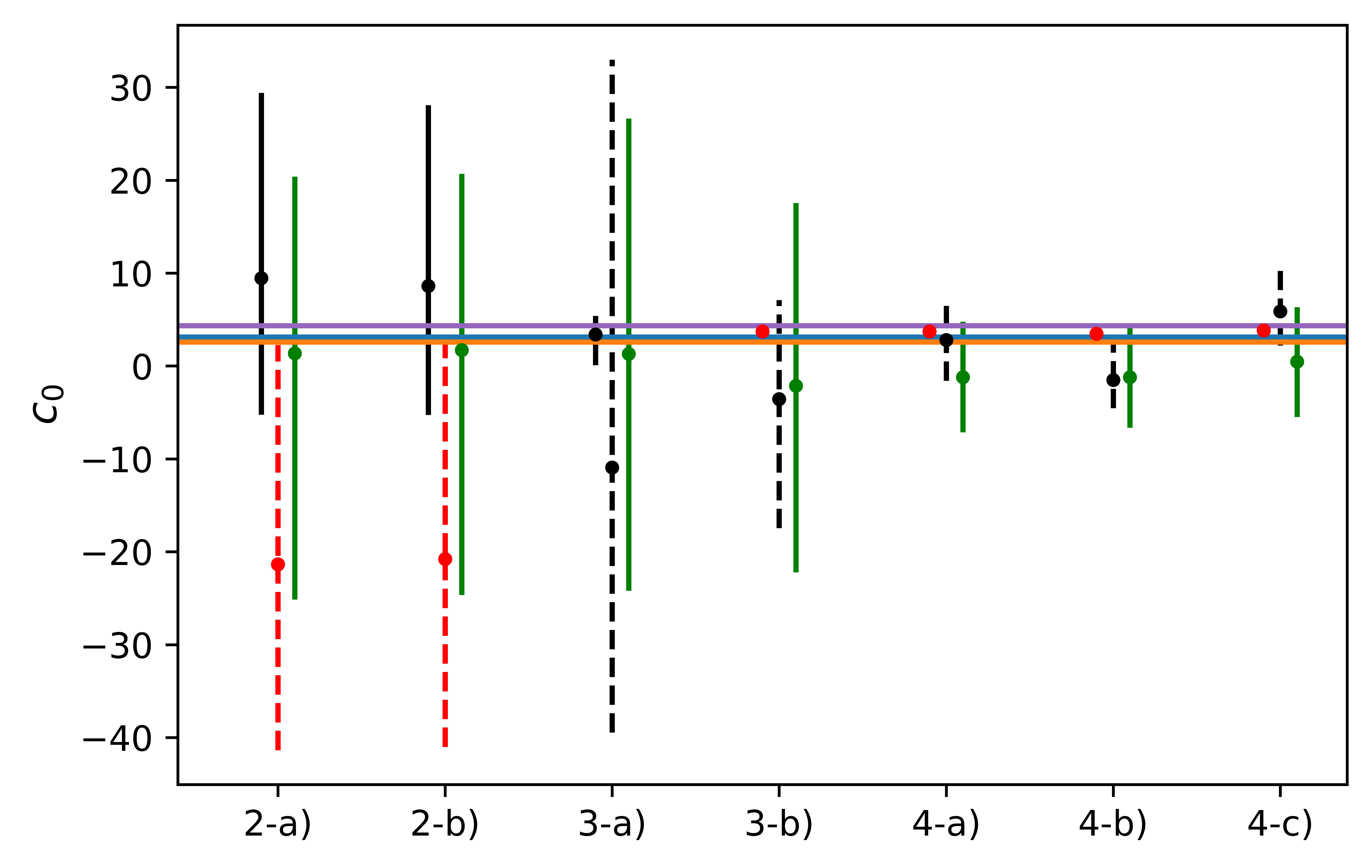}
\includegraphics[width=0.45\columnwidth,keepaspectratio]{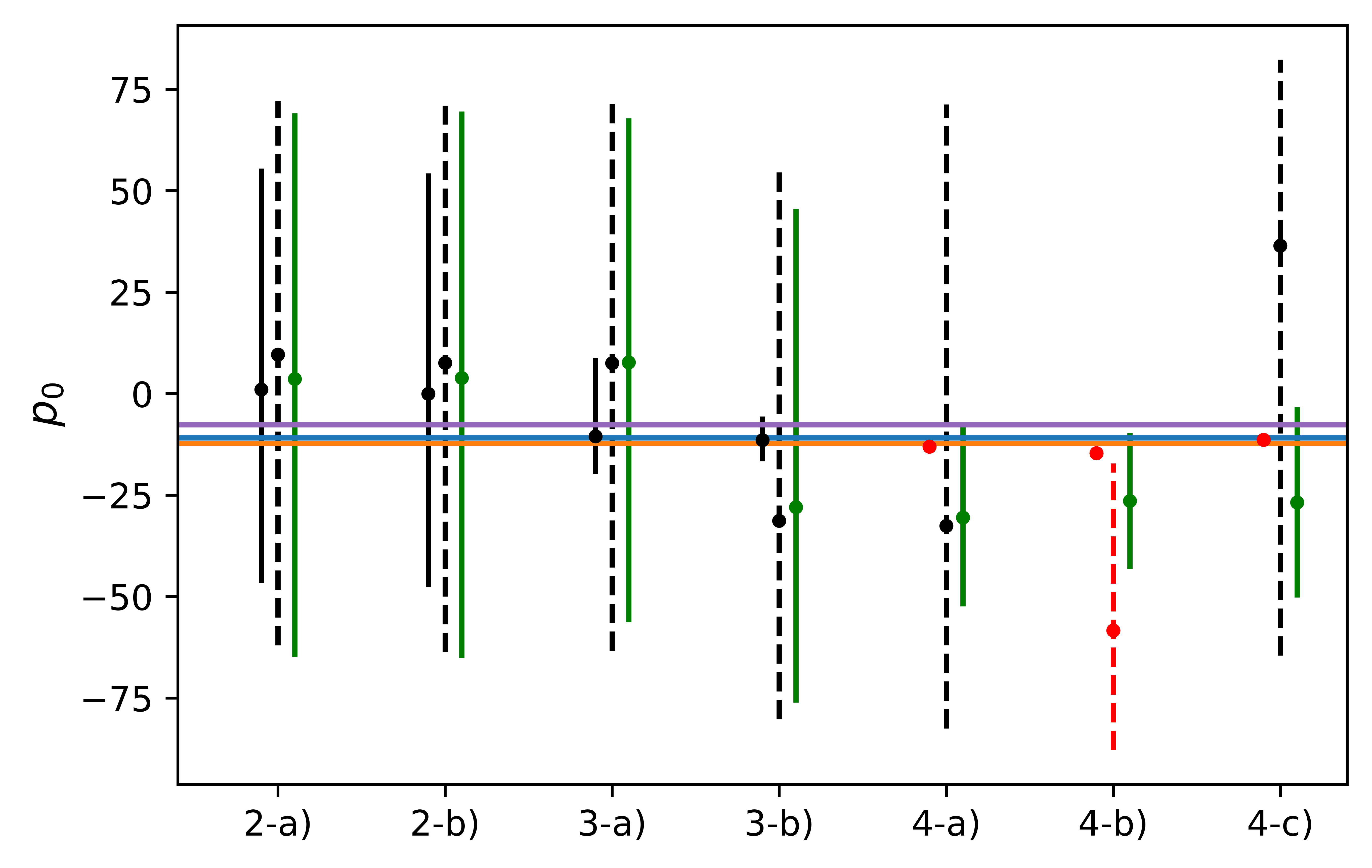}
\end{center}
\caption{A summary of one-sigma derived constraints on the cosmographic coefficients of the three series expansions, presented in Figs. \ref{fig02}--\ref{fig04} and the corresponding tables. The blue, orange. and purple lines identify, the fiducial values of the cosmographic coefficients for the three fiducial models considered: $\Lambda$CDM $(w_0=-1,w_a=0)$, CPL $(w_0=-0.95,w_a=-0.05)$ and CPL $(w_0=-1.0, w_a=0.1)$, respectively, all with $\Omega_m=0.3$ and $H_0=70$ km/s/Mpc. For Figs. \ref{fig02}--\ref{fig03}, $a$ and $b$ denote the left and right panels, while for Fig. \ref{fig04} $a$--$c$ respectively denote the left, middle and right panels. Constraints obtained in the $z$, Pad\'e$[z,3/2]$ and $x$ series are in general shown in black solid, black dashed and green solid error bars, except when the recovered parameter is biased (as defined in the text), in which case it is shown in red.}
\label{fig05}
\end{figure}

We have presented a detailed assessment of the {\bf constraining power and robustness to biases of several possible choices of expansion variables for} the standard cosmographic approach when applied to forthcoming redshift drift measurements by the ELT and the SKAO (in its final configuration). Our analysis relied on simulated data sets representative of both facilities (as well as on an ideal data set), and explored a parameter space of CPL-type cosmological models in the vicinity of the canonical $\Lambda$CDM model. Figure \ref{fig05} provides a visual comparative summary of our results, which have been discussed in the previous section.

Comparing the three standard variables used in the Taylor series ($z$, $y$ and $x$) we find that when describing the redshift dependence of the Hubble parameter the first has the highest mean relative errors at high redshift, making it less reliable in recovering the true values, while the last has the lowest ones. This would suggest that a cosmographic reconstruction based on the $z$ series should be more vulnerable to biases than one based on the $x$ series, which we confirmed in our subsequent analysis. The mean relative errors for the $y$ parameterization are intermediate between the other two, and interestingly (and unlike them) they are independent of every CPL parameter. The use of Pad\'e approximants reduces these biases, but the reduction is largest for the $z$ series and negligible for the $x$ series. 

One of our findings in Sect. \ref{fiducial} is that the spectroscopic velocity, which is the actual observable in the redshift drift case, is more sensitive to the relative errors in the cosmographic approximation, which is even more important considering that one key advantage of redshift drift measurements is the possibility of obtaining high redshift measurements---possibly up to $z=5$. That said, this enhanced sensitivity is not uniform across the various expansion variables, and can still be mitigated through the use of Pad\'e approximants. Relatively speaking the $y$-based expansions are the more vulnerable ones, and the main conclusion is that they are not sufficiently reliable for redshift drift based analyses.

For the low redshift ($z<1$) data that will be provided by the SKAO, the $z$ and $x$ parameterizations can determine the low order cosmographic parameters, up to and including the jerk, with reasonably small uncertainties and without bias. Since $H_0$ is a multiplicative factor in the cosmographic expansion a prior on it is helpful, especially for constraining the deceleration parameter, and a normal prior is advantageous with respect to a uniform one. Pragmatically, we would argue that for low redshift data the $z$ parametrization is adequate, and there is no compelling need to use one of the alternative ones. Therefore this is an opportunity for direct test of  the flat $\Lambda$CDM paradigm, which predicts $j_0=1$.

The Pad\'e$z,[3/2]$ cosmographic case yields nominally small uncertainties but has difficulty in recovering the true values---it other words, it exhibits some bias, even for the jerk parameter. Moreover, in this case one finds peculiarly shaped likelihood distributions, most likely due to the possibility of nuisance asymptotes, which has been reported in the literature \citep{Guthrie}. Indeed we find this to be a generic issue, applying to all the data sets which we have considered, leading to the conclusion that this series lacks numerical reliability to be able to provide reliable reconstructions.

Turning to high-redshift data, as provided by the ELT, the most noteworthy result is how much better the $z$ series does when it comes to constraining the cosmographic coefficients, by comparison to the other two series. This stems from the more complicated dependence of these series on the parameters. This is an issue because higher-order terms in the cosmographic expansion depend on multiple combinations of the cosmographic coefficients, which are prone to degeneracies unless the lower-order terms are well constrained---which generally requires low-redshift data. This is manifest in the fact that, when using ELT data alone, all cosmographic expansions have some difficulty in predicting values for the deceleration parameter, as discussed in the previous section.

The trade-offs between the $z$ and $x$ series can be understood if one considers the combination of the SKAO and ELT data sets. This combination leads to some gains in sensitivity, and these are especially large for the $z$ series, which yields significantly tighter constraints than the other two series, to the extent that even $p_0$ is well constrained in this case. However, this comes at one cost; there is a bias in $c_0$, while the other series (for which the gains mainly occur in the intermediate parameters of the series) are, in this sense, unbiased. In passing, we recall that we have defined a bias to occur when the one-sigma posterior constraint does not recover the parameter's fiducial value.

Finally we have also shown that with ideal data the logarithmic redshift based $x=\ln{(1+z)}$ parametrizations have the most robust behaviour, confirming our earlier result in Sect. \ref{fiducial}, and that with sufficiently precise data redshift drift cosmography can provide discriminating tests between different cosmological models. With such ideal data sets the $z$ parametrization does nominally better in the sense of leading to tighter constraints on the cosmographic coefficients, though usually at the cost of biased in the higher-order ones---in other words, it is only reliable at low redshifts. This opens the possibility of using the redshift drift as a consistency test: one can take a best-fit cosmological model inferred from model-dependent analyses using traditional observables (e.g., Type Ia supernovae, the cosmic microwave background, etc), predict the corresponding cosmographic series, and then use redshift drift measurements to confirm or refute these predictions.

Overall, our conclusion is that for low redshifts the standard $z$ series is perfectly adequate, while for high redshifts the $x$ series provides the best trade-off between variance and bias. The impact of the Pad\'e approximants is somewhat intriguing. In an abstract mathematical sense they are generically expected to be less oscillatory than polynomial functions, from which one would infer that they would be expected to fit a wider range of curves---or, in the present case, data sets. Nevertheless, there will be regions of parameter space where their denominator can become zero, and our analysis indicates that this is a frequent numerical limitation, since there is always some probability that the MCMC will explore those regions. In passing, we note that there are other proposed cosmographic expansions which we have not explored, \textit{e.g.} Chebyshev rationals \citep{Chebyshev} or orthogonalized logarithmic polynomials \citep{Bargiacchi:2021fow}. In future work, it would be interesting to assess how they compare to the $x$ series in the context of redshift drift measurements.

In conclusion, we find that redshift drift cosmography can work well both in the low and high redshift regimes, although the chosen expansion will be important for the recovery of true values, especially in the latter regime. This choice may in part depend on the scientific goals of the analysis and on the sensitivity and redshift range of the available data. As the construction of the ELT and the SKAO progress, realistic end-to-end simulations of these measurements should soon become available (at least in the former case), which will enable more detailed optimization studies. Prospects for a forthcoming reliable model-independent cosmological probe are therefore very good.

\section*{Acknowledgements}

This work was financed by FEDER---Fundo Europeu de Desenvolvimento Regional funds through the COMPETE 2020---Operational Programme for Competitiveness and Internationalisation (POCI), and by Portuguese funds through FCT---Funda\c c\~ao para a Ci\^encia e a Tecnologia in the framework of the project POCI-01-0145-FEDER-028987 and PTDC/FIS-AST/28987/2017.  CJM also acknowledges FCT and POCH/FSE (EC) support through Investigador FCT Contract 2021.01214.CEECIND/CP1658/CT0001.

Several interesting discussions on the topics of this work with Catarina Alves, Axel Lapel, Catarina Marques and Matteo Martinelli are gratefully acknowledged. We also thank the referee, Joe Liske, for the detailed and helpful comments.

{\noindent\bf Data availability:} This work uses simulated data, generated as detailed in the text.

\bibliographystyle{mnras}
\bibliography{drift} 

\begin{thebibliography}{}
\makeatletter
\relax
\def\mn@urlcharsother{\let\do\@makeother \do\$\do\&\do\#\do\^\do\_\do\%\do\~}
\def\mn@doi{\begingroup\mn@urlcharsother \@ifnextchar [ {\mn@doi@}
  {\mn@doi@[]}}
\def\mn@doi@[#1]#2{\def\@tempa{#1}\ifx\@tempa\@empty \href
  {http://dx.doi.org/#2} {doi:#2}\else \href {http://dx.doi.org/#2} {#1}\fi
  \endgroup}
\def\mn@eprint#1#2{\mn@eprint@#1:#2::\@nil}
\def\mn@eprint@arXiv#1{\href {http://arxiv.org/abs/#1} {{\tt arXiv:#1}}}
\def\mn@eprint@dblp#1{\href {http://dblp.uni-trier.de/rec/bibtex/#1.xml}
  {dblp:#1}}
\def\mn@eprint@#1:#2:#3:#4\@nil{\def\@tempa {#1}\def\@tempb {#2}\def\@tempc
  {#3}\ifx \@tempc \@empty \let \@tempc \@tempb \let \@tempb \@tempa \fi \ifx
  \@tempb \@empty \def\@tempb {arXiv}\fi \@ifundefined
  {mn@eprint@\@tempb}{\@tempb:\@tempc}{\expandafter \expandafter \csname
  mn@eprint@\@tempb\endcsname \expandafter{\@tempc}}}

\bibitem[\protect\citeauthoryear{Alves, Leite, Martins, Matos  \& Silva}{Alves
  et~al.}{2019}]{Alves}
Alves C.~S.,  Leite A. C.~O.,  Martins C. J. A.~P.,  Matos J. G.~B.,   Silva
  T.~A.,  2019, \mn@doi [Mon. Not. Roy. Astron. Soc.] {10.1093/mnras/stz1934},
  488, 3607

\bibitem[\protect\citeauthoryear{Balbi \& Quercellini}{Balbi \&
  Quercellini}{2007}]{Balbi}
Balbi A.,  Quercellini C.,  2007, \mn@doi [Mon. Not. Roy. Astron. Soc.]
  {10.1111/j.1365-2966.2007.12407.x}, 382, 1623

\bibitem[\protect\citeauthoryear{Bargiacchi, Risaliti, Benetti, Capozziello,
  Lusso, Saccardi  \& Signorini}{Bargiacchi et~al.}{2021}]{Bargiacchi:2021fow}
Bargiacchi G.,  Risaliti G.,  Benetti M.,  Capozziello S.,  Lusso E.,  Saccardi
  A.,   Signorini M.,  2021, \mn@doi [Astron. Astrophys.]
  {10.1051/0004-6361/202140386}, 649, A65

\bibitem[\protect\citeauthoryear{{Boutsia} et~al.,}{{Boutsia}
  et~al.}{2020}]{Boutsia}
{Boutsia} K.,  et~al., 2020, \mn@doi [\apjs] {10.3847/1538-4365/abafc1}, \href
  {https://ui.adsabs.harvard.edu/abs/2020ApJS..250...26B} {250, 26}

\bibitem[\protect\citeauthoryear{Capozziello, D'Agostino  \&
  Luongo}{Capozziello et~al.}{2018}]{Chebyshev}
Capozziello S.,  D'Agostino R.,   Luongo O.,  2018, \mn@doi [Mon. Not. Roy.
  Astron. Soc.] {10.1093/mnras/sty422}, 476, 3924

\bibitem[\protect\citeauthoryear{Capozziello, D'Agostino  \&
  Luongo}{Capozziello et~al.}{2020}]{Capozziello}
Capozziello S.,  D'Agostino R.,   Luongo O.,  2020, \mn@doi [Mon. Not. Roy.
  Astron. Soc.] {10.1093/mnras/staa871}, 494, 2576

\bibitem[\protect\citeauthoryear{Cattoen \& Visser}{Cattoen \&
  Visser}{2007}]{Cattoen}
Cattoen C.,  Visser M.,  2007, \mn@doi [Class. Quant. Grav.]
  {10.1088/0264-9381/24/23/018}, 24, 5985

\bibitem[\protect\citeauthoryear{Chevallier \& Polarski}{Chevallier \&
  Polarski}{2001}]{CPL1}
Chevallier M.,  Polarski D.,  2001, \mn@doi [Int. J. Mod. Phys.]
  {10.1142/S0218271801000822}, D10, 213

\bibitem[\protect\citeauthoryear{Cooke}{Cooke}{2020}]{Cooke}
Cooke R.,  2020, \mn@doi [Mon. Not. Roy. Astron. Soc.] {10.1093/mnras/stz3465},
  492, 2044

\bibitem[\protect\citeauthoryear{Corasaniti, Huterer  \& Melchiorri}{Corasaniti
  et~al.}{2007}]{Corasaniti}
Corasaniti P.-S.,  Huterer D.,   Melchiorri A.,  2007, \mn@doi [Phys. Rev.]
  {10.1103/PhysRevD.75.062001}, D75, 062001

\bibitem[\protect\citeauthoryear{Darling}{Darling}{2012}]{Darling}
Darling J.,  2012, \mn@doi [Astrophys. J.] {10.1088/2041-8205/761/2/L26}, 761,
  L26

\bibitem[\protect\citeauthoryear{Dong, Gonzalez, Eikenberry, Jeram,
  Likamonsavad, Liske, Stelter  \& Townsend}{Dong et~al.}{2022}]{Dong}
Dong C.,  Gonzalez A.,  Eikenberry S.,  Jeram S.,  Likamonsavad M.,  Liske J.,
  Stelter D.,   Townsend A.,  2022, \mn@doi [Mon. Not. Roy. Astron. Soc.]
  {10.1093/mnras/stac1702}, 514, 5493

\bibitem[\protect\citeauthoryear{Dunsby \& Luongo}{Dunsby \&
  Luongo}{2016}]{Dunsby}
Dunsby P. K.~S.,  Luongo O.,  2016, \mn@doi [Int. J. Geom. Meth. Mod. Phys.]
  {10.1142/S0219887816300026}, 13, 1630002

\bibitem[\protect\citeauthoryear{Esteves, Martins, Pereira  \& Alves}{Esteves
  et~al.}{2021}]{Esteves}
Esteves J.,  Martins C. J. A.~P.,  Pereira B.~G.,   Alves C.~S.,  2021, \mn@doi
  [Mon. Not. Roy. Astron. Soc.] {10.1093/mnrasl/slab102}, 508, L53

\bibitem[\protect\citeauthoryear{Foreman-Mackey}{Foreman-Mackey}{2016}]{corner}
Foreman-Mackey D.,  2016, \mn@doi [The Journal of Open Source Software]
  {10.21105/joss.00024}, 1, 24

\bibitem[\protect\citeauthoryear{Foreman-Mackey, Hogg, Lang  \&
  Goodman}{Foreman-Mackey et~al.}{2013}]{Foreman_Mackey_2013}
Foreman-Mackey D.,  Hogg D.~W.,  Lang D.,   Goodman J.,  2013, \mn@doi
  [Publications of the Astronomical Society of the Pacific] {10.1086/670067},
  125, 306–312

\bibitem[\protect\citeauthoryear{Freedman}{Freedman}{2017}]{Tension}
Freedman W.~L.,  2017, \mn@doi [Nature Astron.] {10.1038/s41550-017-0121}, 1,
  0121

\bibitem[\protect\citeauthoryear{Guthrie}{Guthrie}{2020}]{Guthrie}
Guthrie W.~F.,  2020, NIST/SEMATECH e-Handbook of Statistical Methods (NIST
  Handbook 151), \mn@doi{10.18434/M32189}, \url
  {https://www.itl.nist.gov/div898/handbook/}

\bibitem[\protect\citeauthoryear{Heinesen}{Heinesen}{2021}]{Math2}
Heinesen A.,  2021, \mn@doi [Phys. Rev. D] {10.1103/PhysRevD.104.123527}, 104,
  123527

\bibitem[\protect\citeauthoryear{Klockner et~al.,}{Klockner
  et~al.}{2015}]{Klockner}
Klockner H.-R.,  et~al., 2015, PoS, AASKA14, 027

\bibitem[\protect\citeauthoryear{Linder}{Linder}{2003}]{CPL2}
Linder E.~V.,  2003, \mn@doi [Phys. Rev. Lett.]
  {10.1103/PhysRevLett.90.091301}, 90, 091301

\bibitem[\protect\citeauthoryear{Liske et~al.}{Liske et~al.}{2008}]{Liske}
Liske J.,  et~al., 2008, \mn@doi [Mon. Not. Roy. Astron. Soc.]
  {10.1111/j.1365-2966.2008.13090.x}, 386, 1192

\bibitem[\protect\citeauthoryear{Liske et~al.}{Liske et~al.}{2014}]{HIRES}
Liske J.,  et~al., 2014, Technical report, {Top Level Requirements For
  ELT-HIRES}.
Document ESO 204697 Version 1

\bibitem[\protect\citeauthoryear{Lobo, Mimoso  \& Visser}{Lobo
  et~al.}{2020}]{Math1}
Lobo F. S.~N.,  Mimoso J.~P.,   Visser M.,  2020, \mn@doi [JCAP]
  {10.1088/1475-7516/2020/04/043}, 04, 043

\bibitem[\protect\citeauthoryear{Lu, Jiao, Zhang, Zhang  \& Zhu}{Lu
  et~al.}{2022}]{Lu}
Lu C.-Z.,  Jiao K.,  Zhang T.,  Zhang T.-J.,   Zhu M.,  2022, Phys. Dark Univ.,
  37, 101088

\bibitem[\protect\citeauthoryear{Martinelli, Pandolfi, Martins  \&
  Vielzeuf}{Martinelli et~al.}{2012}]{CMB}
Martinelli M.,  Pandolfi S.,  Martins C. J. A.~P.,   Vielzeuf P.~E.,  2012,
  \mn@doi [Phys. Rev.] {10.1103/PhysRevD.86.123001}, D86, 123001

\bibitem[\protect\citeauthoryear{Martins, Martinelli, Calabrese  \&
  Ramos}{Martins et~al.}{2016}]{Second}
Martins C. J. A.~P.,  Martinelli M.,  Calabrese E.,   Ramos M. P. L.~P.,  2016,
  \mn@doi [Phys. Rev.] {10.1103/PhysRevD.94.043001}, D94, 043001

\bibitem[\protect\citeauthoryear{{McVittie}}{{McVittie}}{1962}]{Mcvittie}
{McVittie} G.~C.,  1962, Ap. J., 136, 334

\bibitem[\protect\citeauthoryear{Moraes \& Polarski}{Moraes \&
  Polarski}{2011}]{Moraes}
Moraes B.,  Polarski D.,  2011, \mn@doi [Phys. Rev.]
  {10.1103/PhysRevD.84.104003}, D84, 104003

\bibitem[\protect\citeauthoryear{Neben \& Turner}{Neben \&
  Turner}{2013}]{Neben}
Neben A.~R.,  Turner M.~S.,  2013, \mn@doi [Astrophys. J.]
  {10.1088/0004-637X/769/2/133}, 769, 133

\bibitem[\protect\citeauthoryear{Press, Teukolsky, Vetterling  \&
  Flannery}{Press et~al.}{2007}]{Press}
Press W.~H.,  Teukolsky S.~A.,  Vetterling W.~T.,   Flannery B.~P.,  2007,
  Numerical Recipes 3rd Edition: The Art of Scientific Computing, 3 edn.
Cambridge University Press, USA

\bibitem[\protect\citeauthoryear{Quercellini, Amendola, Balbi, Cabella  \&
  Quartin}{Quercellini et~al.}{2012}]{Quercellini}
Quercellini C.,  Amendola L.,  Balbi A.,  Cabella P.,   Quartin M.,  2012,
  \mn@doi [Phys. Rept.] {10.1016/j.physrep.2012.09.002}, 521, 95

\bibitem[\protect\citeauthoryear{Rocha}{Rocha}{2021}]{Thesis}
Rocha B. A.~R.,  2021, Master's thesis, University of Porto

\bibitem[\protect\citeauthoryear{{Sandage}}{{Sandage}}{1962}]{Sandage}
{Sandage} A.,  1962, Ap. J., 136, 319

\bibitem[\protect\citeauthoryear{Visser}{Visser}{2005}]{Visser}
Visser M.,  2005, \mn@doi [Gen. Rel. Grav.] {10.1007/s10714-005-0134-8}, 37,
  1541

\bibitem[\protect\citeauthoryear{Xia, Vitagliano, Liberati  \& Viel}{Xia
  et~al.}{2012}]{Xia}
Xia J.-Q.,  Vitagliano V.,  Liberati S.,   Viel M.,  2012, \mn@doi [Phys. Rev.
  D] {10.1103/PhysRevD.85.043520}, 85, 043520

\makeatother
\end{thebibliography}
\appendix

\section{Pad\'e Approximants} 
\label{AppendixA}

For any given power series $A(x)$, the Pad\'e approximant can be calculated using
\begin{equation}
\label{padecalc}
A(x) - \frac{p(x)}{q(x)} = 0\,,
\end{equation}
with $p(x) = p_0+p_1x+p_2x^2+...+p_mx^m$ and $q(x) = 1+q_1x+q_2x^2+...+q_nx^n$. Expanding equation \ref{padecalc} one can determine all the coefficients for $p(x)$ and $q(x)$ as follows
\begin{equation}
\begin{gathered}
\label{system}
a_0 = p_0 \\
a_1 + a_0q_1 = p_1 \\
a_2 + a_1q_1 + a_0q_2 = p_2 \\
... \\
a_m + a_{m-1}q_1 + ... + a_0q_m = p_m \\
a_{m+1} + a_m q_1 + ... + a_{m-n+1}q_n = 0 \\
... \\
a_{m+n} + a_{m+n-1}q_1 + ... + a_m q_n = 0
\end{gathered}
\end{equation}

The numerator ($p(x)$) and denominator ($q(x)$) for the Hubble parameter of the cosmographic Pad\'3[3/2] as a function of redshift $z$ are respectively
\begin{equation}
\begin{split}
\label{zp32p}
& p_{3/2}(z) = H_0\bigg[1+z[(q_0 + 1) + (-13c_0j_0q_0 - 30c_0j_0 + 18c_0q_0^3 + 30c_0q_0^2 + 5c_0s_0 + 130j_0^3q_0 + 120j_0^3 - 280j_0^2q_0^3 \\ & - 230j_0^2q_0^2 + 25j_0^2s_0 - 3j_0p_0 + 270j_0q_0^5 + 240j_0q_0^4 + 40j_0q_0^2s_0 + 110j_0q_0s_0  + 3p_0q_0^2 - 90q_0^7 - 90q_0^6 - 30q_0^3s_0 \\ & + 35q_0s_0^2 + 40s_0^2)/(5(-3c_0j_0 + 3c_0q_0^2 + 12j_0^3 - 23j_0^2q_0^2+ 24j_0q_0^4 + 11j_0q_0s_0 - 9q_0^6 - 3q_0^3s_0 + 4s_0^2))]  \\ & +z^2[(j_0 - q_0^2)/2 + (q_0 + 1)(-13c_0j_0q_0  - 30c_0j_0 + 18c_0q_0^3 + 30c_0q_0^2 + 5c_0s_0  + 130j_0^3q_0 + 120j_0^3 - 280j_0^2q_0^3 \\ & - 230j_0^2q_0^2 + 25j_0^2s_0 - 3j_0p_0  + 270j_0q_0^5 + 240j_0q_0^4 + 40j_0q_0^2s_0 + 110j_0q_0s_0 + 3p_0q_0^2 - 90q_0^7 - 90q_0^6 - 30q_0^3s_0 \\ & + 35q_0s_0^2 + 40s_0^2)/(5(-3c_0j_0 + 3c_0q_0^2 + 12j_0^3 - 23j_0^2q_0^2 + 24j_0q_0^4 + 11j_0q_0s_0 - 9q_0^6 - 3q_0^3s_0 + 4s_0^2)) \\ & + (5c_0^2 - 40c_0j_0^2 + 74c_0j_0q_0^2 - 52c_0j_0q_0 - 60c_0j_0 - 18c_0q_0^4 + 72c_0q_0^3 + 60c_0q_0^2 + 26c_0q_0s_0 + 20c_0s_0 + 80j_0^4 \\ & + 120j_0^3q_0^2 + 520j_0^3q_0 + 240j_0^3 - 475j_0^2q_0^4 - 1120j_0^2q_0^3 - 460j_0^2q_0^2 + 240j_0^2q_0s_0 + 100j_0^2s_0 - 16j_0p_0q_0 - 12j_0p_0 \\ & + 450j_0q_0^6 + 1080j_0q_0^5 + 480j_0q_0^4 - 230j_0q_0^3s_0 + 160j_0q_0^2s_0  + 220j_0q_0s_0 + 60j_0s_0^2 + 12p_0q_0^3 + 12p_0q_0^2 - 4p_0s_0 \\ & - 135q_0^8 - 360q_0^7 - 180q_0^6  + 90q_0^5s_0- 60q_0^3s_0 + 5q_0^2s_0^2 + 140q_0s_0^2 + 80s_0^2)/(20(-3c_0j_0 + 3c_0q_0^2 + 12j_0^3 - 23j_0^2q_0^2  \\ &  + 24j_0q_0^4 + 11j_0q_0s_0 - 9q_0^6 - 3q_0^3s_0 + 4s_0^2))] +z^3[(j_0 - q_0^2)(-13c_0j_0q_0 - 30c_0j_0 + 18c_0q_0^3 + 30c_0q_0^2+ 5c_0s_0  \\ & + 130j_0^3q_0 + 120j_0^3 - 280j_0^2q_0^3 - 230j_0^2q_0^2 + 25j_0^2s_0 - 3j_0p_0 + 270j_0q_0^5 + 240j_0q_0^4  + 40j_0q_0^2s_0 + 110j_0q_0s_0  + 3p_0q_0^2 \\ & - 90q_0^7 - 90q_0^6 - 30q_0^3s_0 + 35q_0s_0^2 + 40s_0^2)/(10(-3c_0j_0  + 3c_0q_0^2 + 12j_0^3 - 23j_0^2q_0^2 + 24j_0q_0^4+ 11j_0q_0s_0 - 9q_0^6  - 3q_0^3s_0 \\ & + 4s_0^2))  + (q_0 + 1)(5c_0^2 - 40c_0j_0^2 + 74c_0j_0q_0^2 - 52c_0j_0q_0 - 60c_0j_0 - 18c_0q_0^4 + 72c_0q_0^3 + 60c_0q_0^2 + 26c_0q_0s_0 \\ & + 20c_0s_0 + 80j_0^4 + 120j_0^3q_0^2 + 520j_0^3q_0 + 240j_0^3 - 475j_0^2q_0^4 - 1120j_0^2q_0^3 - 460j_0^2q_0^2  + 240j_0^2q_0s_0 + 100j_0^2s_0  - 16j_0p_0q_0 \\ & - 12j_0p_0 + 450j_0q_0^6 + 1080j_0q_0^5 + 480j_0q_0^4 - 230j_0q_0^3s_0 + 160j_0q_0^2s_0 + 220j_0q_0s_0 + 60j_0s_0^2 + 12p_0q_0^3 + 12p_0q_0^2 - 4p_0s_0 \\ & - 135q_0^8 - 360q_0^7 - 180q_0^6 + 90q_0^5s_0 - 60q_0^3s_0 + 5q_0^2s_0^2 + 140q_0s_0^2 + 80s_0^2)/(20(-3c_0j_0 + 3c_0q_0^2 + 12j_0^3 - 23j_0^2q_0^2 \\ &  + 24j_0q_0^4 + 11j_0q_0s_0 - 9q_0^6 - 3q_0^3s_0 + 4s_0^2)) - (4j_0q_0 + 3j_0 - 3q_0^3 - 3q_0^2 + s_0)/6]\bigg]
\end{split}
\end{equation}
and
\begin{equation}
\begin{split}
\label{zp32q}
& q_{3/2}(z) = 1+z[(-13c_0j_0q_0 - 30c_0j_0 + 18c_0q_0^3 + 30c_0q_0^2 + 5c_0s_0 + 130j_0^3q_0 + 120j_0^3 - 280j_0^2q_0^3 - 230j_0^2q_0^2 + 25j_0^2s_0 \\ & - 3j_0p_0 + 270j_0q_0^5 + 240j_0q_0^4 + 40j_0q_0^2s_0 + 110j_0q_0s_0 + 3p_0q_0^2 - 90q_0^7 - 90q_0^6 - 30q_0^3s_0 + 35q_0s_0^2 + 40s_0^2)/(5(-3c_0j_0 \\ & + 3c_0q_0^2 + 12j_0^3 - 23j_0^2q_0^2 + 24j_0q_0^4 + 11j_0q_0s_0 - 9q_0^6 - 3q_0^3s_0 + 4s_0^2))]+z^2[(5c_0^2 - 40c_0j_0^2 + 74c_0j_0q_0^2 - 52c_0j_0q_0 \\ &  - 60c_0j_0 - 18c_0q_0^4 + 72c_0q_0^3 + 60c_0q_0^2 + 26c_0q_0s_0 + 20c_0s_0 + 80j_0^4 + 120j_0^3q_0^2 + 520j_0^3q_0 + 240j_0^3 - 475j_0^2q_0^4 - 1120j_0^2q_0^3 \\ & - 460j_0^2q_0^2 + 240j_0^2q_0s_0 + 100j_0^2s_0 - 16j_0p_0q_0 - 12j_0p_0 + 450j_0q_0^6 + 1080j_0q_0^5 + 480j_0q_0^4 - 230j_0q_0^3s_0 + 160j_0q_0^2s_0  \\ & + 220j_0q_0s_0 + 60j_0s_0^2 + 12p_0q_0^3 + 12p_0q_0^2 - 4p_0s_0 - 135q_0^8 - 360q_0^7 - 180q_0^6 + 90q_0^5s_0 - 60q_0^3s_0 + 5q_0^2s_0^2 + 140q_0s_0^2 \\ &  + 80s_0^2)/(20(-3c_0j_0 + 3c_0q_0^2 + 12j_0^3 - 23j_0^2q_0^2 + 24j_0q_0^4 + 11j_0q_0s_0 - 9q_0^6 - 3q_0^3s_0 + 4s_0^2))]\,.
\end{split}
\end{equation}    

On the other hand, for the Hubble parameter of the cosmographic functions Pad\'e[1/4] as a function of the rescaled redshift $y$ the numerator and denominator are respectively
\begin{equation}
\begin{split}
\label{yp14p}
& p_{1/4}(y) = H_0\bigg[1+y[q_0 + 1 + (16c_0q_0 - 230j_0^2q_0 + 735j_0q_0^3 - 200j_0q_0^2 - 200j_0q_0 - 35j_0s_0 + p_0 - 420q_0^5 + 300q_0^4 \\ & + 300q_0^3 + 135q_0^2s_0 - 20q_0s_0 - 20s_0)/(5(c_0 - 10j_0^2 + 105j_0q_0^2 + 40j_0q_0 - 105q_0^4 - 60q_0^3 + 15q_0s_0 + 4s_0))]\bigg]
\end{split}
\end{equation}
and
\begin{equation}
\begin{split}
\label{yp14q}
& q_{1/4}(y) = 1 + y[(16c_0q_0 - 230j_0^2q_0 + 735j_0q_0^3 - 200j_0q_0^2 - 200j_0q_0 - 35j_0s_0 + p_0 - 420q_0^5 + 300q_0^4 + 300q_0^3 \\ &  + 135q_0^2s_0 - 20q_0s_0 - 20s_0)/(5(c_0 - 10j_0^2 + 105j_0q_0^2 + 40j_0q_0 - 105q_0^4  - 60q_0^3 + 15q_0s_0 + 4s_0))] + y^2[(-5c_0j_0 \\ & - 27c_0q_0^2 - 42c_0q_0 - 10c_0 + 50j_0^3 - 115j_0^2q_0^2 + 360j_0^2q_0 + 100j_0^2 - 420j_0q_0^4 - 1620j_0q_0^3 - 650j_0q_0^2 - 5j_0q_0s_0  + 50j_0s_0 \\ & - 2p_0q_0 - 2p_0 + 315q_0^6 + 990q_0^5 + 450q_0^4 - 195q_0^3s_0 - 360q_0^2s_0 - 110q_0s_0)/(10(c_0 - 10j_0^2 + 105j_0q_0^2 + 40j_0q_0 - 105q_0^4 \\ & - 60q_0^3 + 15q_0s_0 + 4s_0))] + y^3[(-13c_0j_0q_0 + 114c_0q_0^3 + 126c_0q_0^2 + 30c_0q_0  + 5c_0s_0 + 340j_0^3q_0 - 300j_0^2q_0^3 + 320j_0^2q_0^2 \\ & + 300j_0^2q_0 + 55j_0^2s_0 - 3j_0p_0 + 1050j_0q_0^5 + 1560j_0q_0^4 + 450j_0q_0^3 + 330j_0q_0^2s_0 + 190j_0q_0s_0 + 60j_0s_0 + 9p_0q_0^2 + 6p_0q_0 \\ & - 630q_0^7 - 1170q_0^6 - 450q_0^5 + 240q_0^4s_0 + 660q_0^3s_0 + 270q_0^2s_0 + 75q_0s_0^2 + 20s_0^2)/(30(c_0 - 10j_0^2 + 105j_0q_0^2 + 40j_0q_0 \\ & - 105q_0^4 - 60q_0^3 + 15q_0s_0 + 4s_0))] + y^4[(-5c_0^2 + 100c_0j_0^2 - 210c_0j_0q_0^2 + 52c_0j_0q_0 + 60c_0j_0 - 210c_0q_0^4 - 456c_0q_0^3 \\ & - 180c_0q_0^2 - 66c_0q_0s_0 - 20c_0s_0 - 500j_0^4 - 700j_0^3q_0^2 - 1360j_0^3q_0 - 600j_0^3 + 1575j_0^2q_0^4 + 1200j_0^2q_0^3 + 100j_0^2q_0^2 - 1020j_0^2q_0s_0  \\ & - 220j_0^2s_0 + 40j_0p_0q_0 + 12j_0p_0 - 3150j_0q_0^6 - 4200j_0q_0^5 - 1200j_0q_0^4 - 210j_0q_0^3s_0 - 1320j_0q_0^2s_0 - 700j_0q_0s_0 - 140j_0s_0^2 \\ & - 60p_0q_0^3 - 36p_0q_0^2 + 4p_0s_0 + 1575q_0^8 + 2520q_0^7 + 900q_0^6 - 630q_0^5s_0 - 960q_0^4s_0 - 300q_0^3s_0 - 285q_0^2s_0^2 - 300q_0s_0^2 \\ & - 80s_0^2)/(120(c_0 - 10j_0^2 + 105j_0q_0^2 + 40j_0q_0 - 105q_0^4 - 60q_0^3 + 15q_0s_0 + 4s_0))]\,.
\end{split}
\end{equation}

Finally, for the Hubble parameter of the cosmographic functions Pad\'e[4/1] as a function of the logarithmic redshift parameter $x$ the numerator and denominator are respectively
\begin{equation}
\begin{split}
\label{xp41p}
& p_{4/1}(x) = H_0\bigg[1 + x[q_0 + 1 + (11c_0q_0 + 5c_0 - 70j_0^2q_0 - 20j_0^2 + 210j_0q_0^3 + 125j_0q_0^2 + 20j_0q_0 - 15j_0s_0 + p_0 - 105q_0^5 \\ & - 75q_0^4 - 15q_0^3 + 60q_0^2s_0 + 35q_0s_0 - q_0 + 5s_0 - 1)/(5c_0 - 20j_0^2  + 125j_0q_0^2 + 40j_0q_0 + 5j_0 - 75q_0^4 - 30q_0^3 - 5q_0^2 \\ & + 35q_0s_0 + 5q_0 + 10s_0 + 5)] + x^2[j_0/2 - q_0^2/2 + q_0/2 + (q_0 + 1)(11c_0q_0 + 5c_0 - 70j_0^2q_0 - 20j_0^2 + 210j_0q_0^3  + 125j_0q_0^2 \\ & + 20j_0q_0 - 15j_0s_0 + p_0 - 105q_0^5 - 75q_0^4 - 15q_0^3 + 60q_0^2s_0 + 35q_0s_0 - q_0 + 5s_0 - 1)/(5c_0 - 20j_0^2 + 125j_0q_0^2 + 40j_0q_0 \\ & + 5j_0 - 75q_0^4 - 30q_0^3 - 5q_0^2 + 35q_0s_0 + 5q_0 + 10s_0 + 5) + 1/2] + x^3[-2j_0q_0/3 + q_0^3/2 + q_0/6 - s_0/6  + (j_0/2 - q_0^2/2 \\ & + q_0/2 + 1/2)(11c_0q_0 + 5c_0  - 70j_0^2q_0 - 20j_0^2 + 210j_0q_0^3 + 125j_0q_0^2 + 20j_0q_0 - 15j_0s_0 + p_0 - 105q_0^5  - 75q_0^4 - 15q_0^3 \\ & + 60q_0^2s_0 + 35q_0s_0 - q_0 + 5s_0 - 1)/(5c_0 - 20j_0^2 + 125j_0q_0^2 + 40j_0q_0 + 5j_0 - 75q_0^4 - 30q_0^3 - 5q_0^2+ 35q_0s_0 + 5q_0  + 10s_0 \\ &  + 5) + 1/6] + x^4[c_0/24 - j_0^2/6 + 25j_0q_0^2/24 + j_0q_0/3 + j_0/24 - 5q_0^4/8 - q_0^3/4 - q_0^2/24 + 7q_0s_0/24 + q_0/24 + s_0/12 \\ & + (-2j_0q_0/3 + q_0^3/2 + q_0/6 - s_0/6 + 1/6)(11c_0q_0 + 5c_0  - 70j_0^2q_0 - 20j_0^2 + 210j_0q_0^3 + 125j_0q_0^2 + 20j_0q_0 - 15j_0s_0 \\ & + p_0 - 105q_0^5 - 75q_0^4 - 15q_0^3 + 60q_0^2s_0 + 35q_0s_0 - q_0 + 5s_0 - 1)/(5c_0 - 20j_0^2 + 125j_0q_0^2 + 40j_0q_0 + 5j_0 \\ & - 75q_0^4 - 30q_0^3 - 5q_0^2 + 35q_0s_0 + 5q_0 + 10s_0 + 5) + 1/24]]\bigg]
\end{split}
\end{equation}
and
\begin{equation}
\begin{split}
\label{xp41q}
& q_{4/1}(x) = 1 + x[(11c_0q_0 + 5c_0 - 70j_0^2q_0 - 20j_0^2 + 210j_0q_0^3 + 125j_0q_0^2 + 20j_0q_0 - 15j_0s_0 + p_0 - 105q_0^5 - 75q_0^4 - 15q_0^3 + 60q_0^2s_0  \\ & + 35q_0s_0 - q_0 + 5s_0 - 1)/(5c_0 - 20j_0^2 + 125j_0q_0^2 + 40j_0q_0  + 5j_0 - 75q_0^4 - 30q_0^3 - 5q_0^2 + 35q_0s_0 + 5q_0 + 10s_0 + 5)]
\end{split}
\end{equation}

\bsp	
\label{lastpage}
\end{document}